\documentclass[a4paper,fleqn,usenatbib]{mnras}

\usepackage[T1]{fontenc}
\usepackage{ae,aecompl}

\usepackage{graphicx}	
\usepackage{amsmath}	
\usepackage{amssymb}	
\usepackage{pdflscape}
\usepackage{threeparttable}
\usepackage{ulem}
\usepackage{booktabs}
\usepackage{natbib}

\title[Three new barium dwarfs with WD companions]{Three new barium dwarfs with WD companions: BD+68$^\circ$1027, RE~J0702+129 and BD+80$^\circ$670}

\author[X. M. Kong et al.]{
 X. M. Kong,$^{1,2,3}$\thanks{Contact e-mail: \href{mailto:xmkong@nao.cas.cn}{xmkong@nao.cas.cn}}
 Y. Bharat Kumar ,$^{1}$
 G. Zhao,$^{1,3}$\thanks{Contact e-mail: \href{mailto:gzhao@nao.cas.cn}{gzhao@nao.cas.cn}}
 J. K. Zhao,$^{1}$ X.S. Fang, $^{1}$
 \newauthor J. R. Shi, $^{1,3}$
 L. Wang, $^{1}$
 J. B. Zhang, $^{1}$
 and H. L. Yan  $^{1}$
 \\ \\
 $^{1}$ Key Laboratory of Optical Astronomy, National Astronomical Observatories, Chinese Academy of Sciences, Beijing 100012, China \\
 $^{2}$ School of Mechanical, Electrical and Information Engineering, Shandong University at Weihai, Weihai 264209, China \\
 $^{3}$ School of Astronomy and Space Science, University of Chinese Academy of Sciences, Beijing 100049, China
 }
 \date{Accepted XXX. Received YYY; in original form ZZZ}

\pubyear{2017}

\begin{document}
\label{firstpage}
\pagerange{\pageref{firstpage}--\pageref{lastpage}}
\maketitle

\begin{abstract}
We report three new barium (Ba) dwarfs lying in Sirius-like systems, which provides direct evidence that Ba dwarfs are companions to white dwarfs (WDs).
Atmospheric parameters, stellar masses, and chemical abundances of 25 elements, including light, $\alpha$, Fe-peak and s-process elements, are derived from high resolution and high S/N spectra. Enhancement of s-process elements with [s/Fe] ratios between 0.4 and 0.6 confirm them as mild barium stars. The estimated metallicities ($-$0.31, $-$0.06, 0.13) of BD+68$^\circ$1027, RE~J0702+129 and BD+80$^\circ$670 are in the range of known Ba dwarfs and giants. As expected, observed indices of [hs/ls], [s/Fe] and [C/Fe] show anticorrelation with metallicity.
AGB progenitor masses are estimated for the WD companions of RE~J0702+129 (1.47 $M_{\odot}$) and BD+80$^\circ$670 (3.59 $M_{\odot}$), which confirms the predicted range of progenitor AGB masses (1.5 $\sim$ 4 $M_{\odot}$) for unseen WDs around Ba dwarfs.
 Surface abundances of s-process elements in RE~J0702+129 and BD+80$^\circ$670 are compared with AGB models and they are in close agreement, within predicted accretion efficiencies and pollution factors for Ba stars.
These results support that the origin of s-process overabundances in Ba dwarfs is similar to Ba giants via McClure hypothesis in which Ba stars accumulate s-process elements through mass transfer from their host companions during AGB phase.
\end{abstract}




\section{Introduction}

Classical Ba stars are GK-type giant stars that show enhancement in carbon and heavy elements produced by the s-process nucleosynthesis, firstly discovered by \cite{Bidelman1951}. The production of neutron capture elements in a star is expected to happen on the thermally pulsating phase of asymptotic giant branch (TP-AGB), but the luminosity estimations for Ba stars are below the threshold for the onset of TP-AGB to experience self-enrichment of s-process elements \citep{1997Bergeat, 2017Escorza}.
\citet{McClure1980} proposed a mass-transfer scenario within a binary system with a possible companion WD (AGB progenitor) that could be responsible for s-process overabundances in these stars.

In the last few decades, extensive studies were performed for checking binarity in Ba stars \citep{McClure1984, 1990McClure, Jorissen1998, Vitense2000}. For example, \cite{Vitense2000} searched for properties of companions based on characteristic ultraviolet continua in several Ba stars and found definite excess fluxes that can be attributed to WD companions for five of their targets, and concluded that it is indeed highly probable that all barium and mild barium stars have WD companions. It was quite recently, based on near-UV and far-UV fluxes data from Galaxy Evolution Explorer (GALEX mission), \cite{Gray2011} quantified far-UV excess in six Ba dwarfs compared to normal dwarfs and presented a direct argument that Ba dwarfs have WD companions. These results basically supported the binary origin for these Ba stars, although negative cases might exist, such as those identified by Escorza (private communication) from a long-term radial-velocity monitoring of five Ba dwarfs uncovered by \cite{1993Edvardsson}. They found that two of them are not binaries. On the other hand, it seems that a binary system with a WD companion is not a sufficient condition for a star to become a barium star. \cite{Merle2016} analysed the abundances of s-process elements for primary components of 11 binary stars and concluded that binary systems with WD companions less massive than 0.5 ${M_\odot}$ do not form barium stars.

The external contamination hypothesis predicts that the star contaminated by its AGB companion is as likely to be a dwarf as a giant \citep{Gray2011}. Moreover, the WD would cool off after the mass transfer, and many WD companions are so cool that their cooling timescales must be longer than the lifetime of barium stars on the giant branch \citep{Vitense2000}. So, a fairly large number of main-sequence stars with s-process overabundances are expected. But the fact is, Ba stars are known to be associated with giant branch since 1951 until the discovery of mild s-process overabundances in a dwarf star HR~107 was reported by \cite{Tomkin1989}, which suggests the existence of main-sequence counterparts of classical Ba giants. Following the discovery of the Ba \mbox{dwarf} star HR~107, \cite{1993Edvardsson} discovered 5 Ba \mbox{dwarfs} from the large survey \mbox{meant} for abundance study of F \mbox{stars}.
\cite{1994North} discovered 8 Ba dwarfs among 20 F stars with strong \mbox{Sr} $\lambda4077$ mentioned by \cite{1981Bidelman, 1983Bidelman, 1985Bidelman} and analyzed three Ba dwarfs taken from the list of \cite{1983Lu} which are probably dwarfs rather than giants. In the next twenty years, dedicated surveys increased the number of Ba dwarfs (HD~147513, \citealt{1997Porto};  HD~8270, HD~13551, HD~22589, \citealt{2005Pereira}; HD~26367, \citealt{2007Gray}; HD~11397, HD~14282, \citealt{2008Pomp}; BD$-$03$^\circ$3668, \cite{Pereira2011}). But compared to Ba giants, the sample of Ba dwarfs known to date is considerably smaller. It is especially \mbox{important} to identify new examples of Ba dwarfs and provide more direct evidences that Ba dwarfs have WD companions, for investigating the origin and properties of Ba stars.

Sirius-like systems (SLSs) are binary systems wherein dwarfs and giants are companions with WDs, and some primaries are found to be Ba giants.
\cite{Holberg2013} catalogued 98 SLSs including the information on orbital periods and WD companion masses. To search for Ba dwarfs and investigate the origin and properties of Ba stars, we chose 21 primary stars in SLSs, including 6 giants and 15 dwarfs.
We identified three Ba dwarfs (BD+68$^\circ$1027, BD+80$^\circ$670 and RE~J0702+129) among our sample based on their detailed abundance analysis, and their abundance patterns are compared with known barium dwarfs, CH subgiants, and AGB models in the present work.

In the following Section 2, we describe the observations and process of data reduction. Determination of stellar parameters are presented in Section 3. Determination of abundances and their uncertainties are presented in Section 4. Comments on individual stars are given in Section 5. The results are discussed in Section 6 and concluded in Section 7.

\section{High-Resolution spectra of the sample stars}
 Sample stars are selected from \citet{Holberg2013} and their basic parameters are given in Table \ref{tab:BasInf}.
For the stars which have parallaxes from both Hipparcos and Gaia DR1 TGAS data \citep{2016Gaia}, we adopted the distances from Gaia to calculate relevant atmospheric parameters.

\begin{table*}

\centering
\caption{ Log of observations and basic data of sample stars.    }
\label{tab:BasInf}
\scalebox{0.9}{
\begin{tabular}{l|ccccccccccc}
\hline\hline
Primary&
Date &
Exposure time &
Primary&
 {V}  &
$ {\varpi}${(Hip)}    &
d    &
$ {\varpi}${(Gaia)} &
d    &
WD   &
$M_\mathrm{WD}$
 \\
  &
  &
 (sec)   &
 Type  &
   &
 {(mas)} &
(pc)  &
 {(mas)} &
(pc)   &
 Type  &
$({M_\odot})$ &

   \\

\hline
BD+68$^\circ$1027	&	2014 Nov. 11	&	1800	&	G5	    &	9.78 	&	$12.68 \pm 0.76$	&	78.86 	   &	$11.09 \pm 0.22$ &	90.17 	&	DA	    &	-  \\
RE~J0702+129	&	2014 Oct. 23	&	1200	&	K0IV/V	&	10.66 	&	-                   &  115.00$^a$ 	&	-	              & -	    &	DA1.4	&	0.57$^b$   \\
BD+80$^\circ$670	&	2014 Oct. 23	&	600	   &	G5V	    &	9.16 	&  -                    &	40.00$^c$ 	       &	$11.67 \pm 0.72$  &  85.69 	&	DA6	&	0.81$^c$  \\
\hline

\multicolumn{11}{l}{$a$ \cite{2010Barstow};  $b$ \cite{2010Kawka};  $c$ \cite{2011Gianninas}}
\end{tabular}
}
\end{table*}

 \subsection{Observations and data reduction}
\begin{figure}

	\includegraphics[width=\columnwidth]{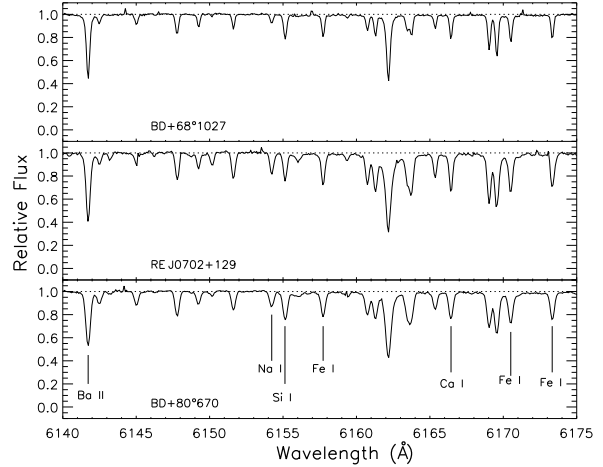}
    \caption{A portion of spectra of sample stars in the wavelength region of 6140 to 6175 \AA. Some lines are identified.}
    \label{fig:spec}

\end{figure}

The spectra of three sample stars were obtained using ARC Echelle Spectrograph (ARES) mounted on 3.5 m telescope located at Apache Point Observatory (APO), during two observing runs in 2014 (see Table \ref{tab:BasInf} for details). Spectra cover wavelength range of 4,400\,{\AA} to 10,000\,{\AA} with a resolution $R\sim31,500\,$.
The log of observations is given in Table~\ref{tab:BasInf} and sample spectra in Ba region are shown in Figure~\ref{fig:spec}. For the three stars, S/N ratio is over 100 in the entire wavelength region. In addition, we obtained the solar spectrum for performing differential abundance analysis.

The raw 2D spectra were reduced using IDL \mbox{programs} for background subtraction, flat-fielding, order identification, extraction, and wavelength calibration. The 1-D wavelength calibrated spectra were normalized to continuum by fitting cubic spline functions with smooth \mbox{parameters}. Radial velocities were obtained by fitting the profiles of about 80 pre-selected lines with inter-mediate strength (See \citealt{WANG2011} for details).

\subsection{Equivalent widths}
We measured the equivalent widths (EWs) in continuum normalized spectra using two methods: for weak lines, the line profiles were usually well fitted by a Gaussian function, and direct integration was adopted for unblended lines which are well separated from nearby lines.
For most of the elements, we chose lines with EWs between 10 and 120 m{\AA}. For some elements (e.g. K, and Ba), we opted to use lines with EWs stronger than 120 m{\AA} due to limited availability of spectral lines throughout the wavelength region of 4,500-9,000 {\AA}.

\begin{figure}

	\includegraphics[width=\columnwidth]{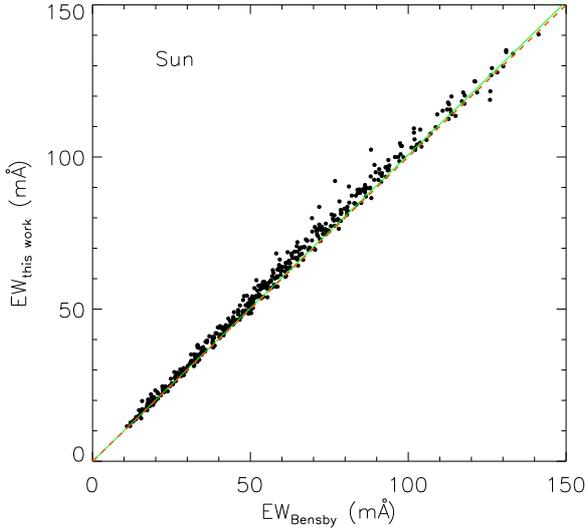}
    \caption[Cap for listoffigures]{A comparison of equivalent widths measured in this work with those of \cite{Bensby2014} for Sun. The solid line is a linear fit to the points, whereas the dashed line is one-to-one correlation.}
    \label{fig:ew}

\end{figure}

The uncertainties of EWs were estimated by comparing them for the solar spectrum with the values taken from \cite{Bensby2014}.
The comparison of EWs for 452 lines is displayed in Figure \ref{fig:ew}. We obtained a linear regression for the two data sets, $EW_\mathrm{this\,work}$ = 1.007 $(\pm 0.003)$ $EW_\mathrm{Bensby}$ - 0.023 $(\pm0.021)$ {(m\AA)}, and the standard deviation is $\sim$2.5 m{\AA}.

\section{Stellar atmospheric parameters}
\label{sec:atmospheric} 

\subsection{Effective temperature}

\begin{figure}

	\includegraphics[width=\columnwidth]{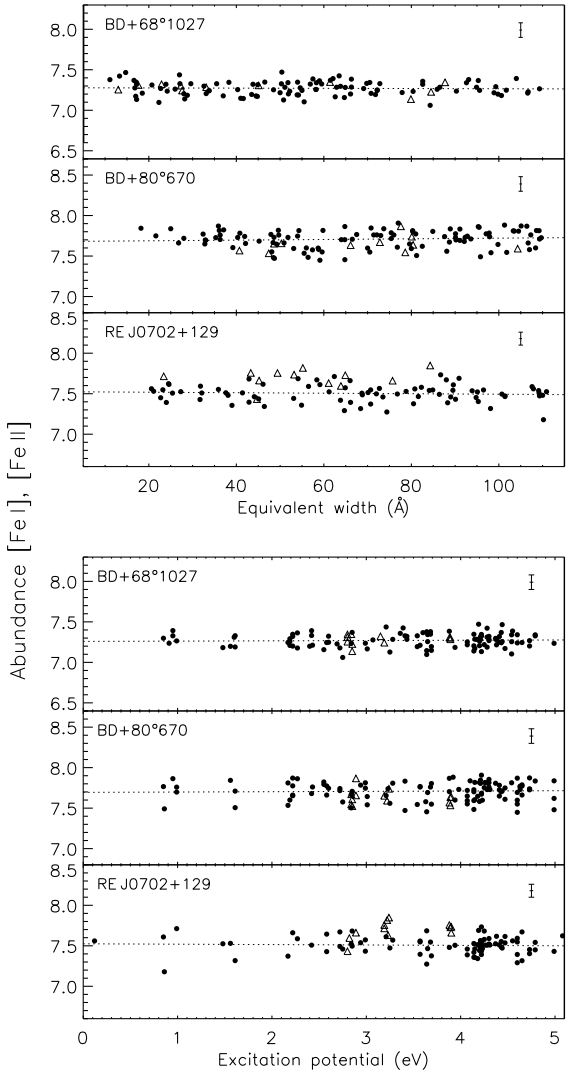}
    \caption{Iron abundances from Fe\,{\sc i} (filled circles) and Fe\,{\sc ii} (open triangles) lines are  shown as function of equivalent widths (top) and lower excitation potentials (bottom). The typical error bars are shown in the top right-hand corner of each panel, they are also listed in Table \ref{tbl:error1}.
    }
    \label{fig:teff}

\end{figure}

The effective temperature ($T_\mathrm{eff}$) was determined from excitation equilibrium method, which requires Fe\,{\sc i} spectral lines with a range of lower excitation potentials to give \mbox{equal} abundances. The slopes of $\log A$ - $\chi_{\mathrm{low}}$ diagrams for our sample stars are smaller than 0.005 dex/eV.
The abundances of Fe\,{\sc i} and Fe\,{\sc ii} of the three stars as a function of excitation potentials and equivalent widths are shown in Figure~\ref{fig:teff}.
For comparison, we used the photometric colour index ($V-K$) and empirical calibration relations given by \cite{Alonso1996} to derive effective temperatures. $J$ and $K_\mathrm{s}$ magnitudes were taken from 2MASS \citep{2003Cutri} and converted to $K$ magnitude (Telescopio Carlos S\'{a}nchez photometric system) using the calibration relation given by \cite{Ramirez2004}. $V$ magnitudes were obtained by converting $B_\mathrm{t}$ and $V_\mathrm{t}$ magnitudes from $Tycho$ to $Johnson$ $V$ system using the calibration relation given by \cite{Mamajek2002}.
The color excess $E(B-V)$ was interpolated from the dust maps from \cite{Schlegel1998}, with a slight revision (see \cite{2002Beers} for details). Then, we adopted $E(V-K) = 2.727 E(B-V)$ as the colour excess for $V-K$ \citep{McCall2004}. Table \ref{para} lists the effective temperatures derived from $V-K$ and excitation equilibrium method, and we adopted the latter as our final effective temperature in following analysis.

For RE~J0702+129, due to lack of parallax from Hipparcos or Gaia, we adopted distance of 115 pc from \cite{2010Barstow} to estimate reddening and effective temperature, $T_\mathrm{eff}(V - K)$ = 5052 K. The difference between photometric and excitation equilibrium temperature ($T_\mathrm{eff}(Spec)$ = 5531 K), is found to be 479 K. In addition, using $T_\mathrm{eff}$ = $T_\mathrm{eff}(V - K)$ of 5052 K we found the difference in iron abundances between Fe\,{\sc i} and Fe\,{\sc ii} is over 0.5 dex (A(Fe\,{\sc i}) = 7.28, A(Fe\,{\sc ii}) = 7.82). Photometric temperature from $V - K$ may be under estimated due to the presence of high level activity in this star (see Sec. 5.1) that could affect the color index \citep{2016Covey}. Another reason could be that RE~J0702+129 is in a triple system and the third companion might have influenced the observed V or K magnitude.

\subsection{Surface gravity}

The surface gravity ($\log g$) was determined using two methods. One is so-called ionization equilibrium method (spectroscopic)  wherein the value of $\log g$ is derived from ionization balance between Fe\,{\sc i} and Fe\,{\sc ii}, and the other is from basic principles through the relationship between bolometric flux, temperature, mass and gravity. For the latter, the fundamental relation is :

\begin{equation}
\label{math:logg}
\log{g} = \log{g_\odot} +
          \log\left(\frac{M}{M_\odot}\right) +
          4\log\left(\frac{T_\mathrm{eff}}{T_{\mathrm{eff}\odot}}\right) +
          0.4(M_{\mathrm{bol}}-M_{\mathrm{bol}\odot})
\end{equation}

\begin{equation}
\label{math:Mbol}
    M_{\mathrm{bol}}=V_\mathrm{mag}+BC+5\log{\varpi}+5-A_{\mathrm{V}}
\end{equation}

where $M$ is the stellar mass, which is estimated using an interpolator of the evolutionary tracks of \cite{YonseiYale2003}. $V_\mathrm{mag}$, $\varpi$, $BC$, $M_\mathrm{bol}$ and $A_{\mathrm{V}}$ represent apparent magnitude, parallax, bolometric correction, absolute bolometric magnitude and interstellar extinction, respectively.
The bolometric corrections were calculated using the relation given by \cite{Alonso1995}, which depends on temperature and metallicity.
For BD+68$^\circ$1027 and BD+80$^\circ$670, we derived $\log g$ by two methods and adopted the one derived from Gaia parallaxes.
Table \ref{para} list the surface gravities of sample stars derived from both methods.

\subsection{Microturbulent velocity and metallicity}
Microturbulence ($\xi_{t}$) was determined by forcing the iron abundances from different Fe\,{\sc i} lines to be independent from their EWs. Only those Fe\,{\sc i} lines with $10$ m{\AA} $ < EW <110$ m{\AA} were selected in all three sample stars.
 The initial metallicity values for our program stars were set to [Fe/H]= 0.0.

 We adopted the final results by iterating the whole processes of determining the atmospheric parameters $T_{\rm eff}$, $\log g$, [Fe/H], and $\xi_{t}$ until they were consistent.

\begin{table*}
\centering
\caption{ Stellar parameters of the program stars.    }
\label{para}
\begin{tabular}{l|ccrcccccrr}
\hline\hline
Primary  &
V-K       &
E(V-K)    &
$M_\mathrm{v}$    &
$M_\mathrm{p}/{M_\odot}$    &
$T_\mathrm{eff}$  &
$T_\mathrm{eff}$  &
$\log g$       &
$\log g$        &
$\mathrm{[Fe/H]}$  &
$\xi_\mathrm{t}$
 \\
 &
 &
 &
 &
 &
 (V-K)&
 (Spec)&
 (Parallax)&
 (Spec)  \\

\hline

BD+68$^\circ$1027	&	1.49 	&	0.04 	&	5.01 	&	0.93 	&	5819	&	5919	&	4.48	&	4.55 	&	$-$0.31	&	1.0 	\\
RE~J0702+129	&	2.06 	&	0.03 	&	5.28	&	0.93	&	5052	&	5531	&	4.43	&	4.20 	&	$-$0.06	&	1.9 	\\
BD+80$^\circ$670	&	1.57 	&	0.13 	&	4.50 	&	1.05	&	5840	&	5880	&	4.33	&	4.50 	&	0.13	&	1.6 	\\

\hline
\end{tabular}

\end{table*}

\begin{table} \small
\caption{Atomic line data with their equivalent widths and abundances of the program stars.
    }
\label{tbl:linlst}
\centering
\scalebox{0.7}{
\begin{tabular}{c|clcccrr}
\hline\hline
Star   &
Wavelength  &
Elem  &
Ion &
$\log gf$           &
LEP  &
EW  &
Abun \\
    &
[{\AA}]  &
    &
&    & [eV] &(m{\AA})  & (dex)  \\
 \hline
 BD+68$^\circ$1027	&	5380.337 	&	C	&	\,{\sc i}	&	7.680 	&	$-$1.57	&	16.1	&	8.22 	\\
	&	5052.170 	&	C	&	\,{\sc i}	&	7.680 	&	$-$1.24	&	27.2	&	8.17 	\\
	&	5688.217 	&	Na	&	\,{\sc i}	&	2.100 	&	$-$0.42	&	104.8	&	6.18 	\\
	&	6154.230 	&	Na	&	\,{\sc i}	&	2.100 	&	$-$1.51	&	18.4	&	5.95 	\\
	&	6160.753 	&	Na	&	\,{\sc i}	&	2.100 	&	$-$1.24	&	36.7	&	6.07 	\\

\hline
\end{tabular}
}
\end{table}

\section{Abundance analysis}

\subsection{Atomic data}
The atomic data for most of Fe\,{\sc i} lines used in this \mbox{study} were taken from \cite{Bensby2014}. To increase the accuracy of $T_{\rm eff}$ (derived from excitation equilibrium method) and $\log g$ (determined from ionization equilibrium method), we selected 30 weak Fe\,{\sc i} lines and 10 Fe\,{\sc ii} lines from \cite{Blackwell1982b, Blackwell1982c}, \cite{OBrian1991} and \cite{BardKock1991} or \cite{Bard1994} as a supplement.
The references for atomic data of other elements are following: O\,{\sc i} \citep{2007Liu}, Ce\,{\sc ii}, Y\,{\sc ii} \citep{Allen2011}, La\,{\sc ii}, Zr\,{\sc ii} \citep{2013Johnson}, Co\,{\sc i}, \citep{Colucci2012}, Nd\,{\sc ii} \citep{2011Pereira}, C\,{\sc i}, Cu\,{\sc i}, Sc\,{\sc i}, Sc\,{\sc ii}, V\,{\sc i}, Mn\,{\sc i}, K\,{\sc i}, Sr\,{\sc i} \citep{Melendez2014}, Ba\,{\sc ii}, Ca\,{\sc i}, Cr\,{\sc i}, Cr\,{\sc ii}, Ni\,{\sc i}, Ti\,{\sc i}, Ti\,{\sc ii}, Si\,{\sc i}, Mg\,{\sc i}, Na\,{\sc i} \citep{Bensby2014}.

In order to inspect the quality of these $gf$ values, we obtained abundances of the Sun using our solar spectrum, and the agreement with those given by \cite{Grevesse1998} was taken as an evidence for reliability of selected atomic line data. During this process, the deviation between the abundance derived from given lines and the mean abundance from all lines of the same element was examined, and we discarded the lines with large deviation($>$0.25 dex). Finally, $T_{\rm eff}$, $\log g$ and $\xi_{t}$ for the sun were adopted as 5780 K, 4.44, and 0.9 km\,s$^{-1}$, respectively, and the standard deviation of each element abundance is within 0.11 dex.
The EWs and derived elemental abundances for our program \mbox{stars along} with atomic data are listed in Table \ref{tbl:linlst}. Here we show few lines for guidance and the complete table is available in electronic form at the CDS.

\begin{table*} \small
\caption{Derived differential abundances for 25 elements in the program stars.}
\label{tbl:abundance}
\centering

\begin{tabular}{l|rrrrrr}
\hline\hline

     &

 BD+68$^\circ$1027 &
  &
 BD+80$^\circ$670  &
    &
RE~J0702+129          &

\\

Species  &
[X/Fe] &
No.Lines    &
   [X/Fe]  &
No.Lines &
[X/Fe]      &
No.Lines     \\

\hline
Li	&	        -   	&	- 	&	     -	        &	- 	&	$1.55\pm0.05$ (A(Li))	&	1 	\\
  C	&	$	0.11 	\pm	0.04 	$	&	2	&	$	-0.13 	\pm	0.00 	$	&	2	&	$	0.16 	\pm	0.04 	$	&	2	\\
O	&	$	0.07 	\pm	0.08 	$	&	3	&	$	-0.20 	\pm	0.08 	$	&	2	&	$	0.17 	\pm	0.03 	$	&	2	\\
  Na	&	$	0.00 	\pm	0.08 	$	&	3	&	$	0.04 	\pm	0.03 	$	&	2	&	$	0.12 	\pm	0.03 	$	&	2	\\
  Mg	&	$	-0.03 	\pm	0.04 	$	&	5	&	$	-0.03 	\pm	0.04 	$	&	2	&	$	-0.06 	\pm	0.01 	$	&	3	\\
  Al	&	$	0.03 	\pm	0.03 	$	&	3	&	$	0.08 	\pm	0.06 	$	&	4	&	$	0.15 	\pm	0.03 	$	&	3	\\
  Si	&	$	0.03 	\pm	0.07 	$	&	17	&	$	-0.02 	\pm	0.06 	$	&	16	&	$	-0.12 	\pm	0.07 	$	&	9	\\
   K	&		0.15 				&	1	&	$	0.18 			$	&	1	&	$	0.54 			$	&	1	\\
  Ca	&	$	0.09 	\pm	0.07 	$	&	11	&	$	0.09 	\pm	0.05 	$	&	8	&	$	0.15 	\pm	0.02 	$	&	5	\\
 Sc	&	$	-0.03 	\pm	0.09 	$	&	4	&	$	-0.14 	\pm	0.05 	$	&	3	&	$	-0.15 	\pm	0.04 	$	&	3	\\
 TiI	&	$	0.04 	\pm	0.06 	$	&	14	&	$	-0.05 	\pm	0.07 	$	&	14	&	$	-0.01 	\pm	0.09 	$	&	6	\\
TiII	&	$	-0.06 	\pm	0.08 	$	&	5	&	$	-0.18 	\pm	0.07 	$	&	5	&	$	-0.05 	\pm	0.10 	$	&	6	\\
   V	&	$	-0.13 	\pm	0.04 	$	&	2	&	$	0.05 	\pm	0.05 	$	&	4	&	$	0.20 	\pm	0.02	$	&	2	\\
 CrI	&	$	0.04 	\pm	0.06 	$	&	5	&	$	0.06 	\pm	0.02 	$	&	5	&	$	0.17 	\pm	0.01 	$	&	2	\\
CrII	&	$	0.01 	\pm	0.03 	$	&	6	&	$	-0.11 	\pm	0.04 	$	&	4	&	$	0.07 	\pm	0.02 	$	&	3	\\
  Mn	&	$	-0.20 	\pm	0.10 	$	&	4	&	$	-0.02 	\pm	0.03 	$	&	4	&	$	-0.06 			$	&	1	\\
FeI	&	$	-0.31 	\pm	0.11 	$	&	112	&	$	0.13 	\pm	0.11 	$	&	112	&	$	-0.06 	\pm	0.11 	$	&	84	\\
FeII	&	$	-0.34 	\pm	0.06 	$	&	11	&	$	0.01 	\pm	0.09 	$	&	13	&	$	-0.06 	\pm	0.11 	$	&	12	\\
  Co	&	$	0.00 			$	&	1	&		$-0.10 \pm 0.09	$		&	3	&	$	-0.16			$	&	1	\\
 Ni	&	$	-0.03 	\pm	0.08 	$	&	25	&	$	-0.05 	\pm	0.08 	$	&	30	&	$	-0.07 	\pm	0.09 	$	&	23	\\
  Cu	&	$	-0.14 	\pm	0.06 	$	&	2	&	$	-0.09 			$	&	1	&	$	-0.24 			$	&	1	\\
  Sr	&	$	0.58 			$	&	1	&	$	0.96 			$	&	1	&	$	0.65 			$	&	1	\\
   Y	&	$	0.51 	\pm	0.07 	$	&	3	&	$	0.46 	\pm	0.06 	$	&	2	&	$	0.33 	\pm	0.02 	$	&	2	\\
 Zr	&	$	0.46 			$	&	1	&	$	0.38 			$	&	1	&	$	0.38 			$	&	1	\\
  Ba	&	$	0.62 	\pm	0.07 	$	&	3	&	$	0.31 	\pm	0.03 	$	&	3	&	$	0.56 	\pm	0.06 	$	&	2	\\
  La	&	$	0.57 			$	&	1	&		-				&	-	&	$	0.53 			$	&	1	\\
  Ce	&	$	0.72 	\pm	0.07 	$	&	3	&	$	0.25 			$	&	1	&	$	0.25 	\pm	0.00 	$	&	2	\\
  Nd	&	$	0.55 	\pm	0.04 	$	&	2	&		-				&	-	&	$	0.13 	\pm	0.10 	$	&	2	\\

\hline
\end{tabular}

\end{table*}

\subsection{Abundances and their uncertainties}

Elemental abundances were determined based on measured EWs of spectral lines with reliable atomic data, model atmospheres, and ABONTEST8 program.
The required model atmospheres are interpolated from a grid of plane-parallel, local thermodynamic equilibrium model provided by \cite{1993Kur}. The ABONTEST8 program supplied by Dr. P. Magain was used to calculate the theoretical line EWs, and final abundances were obtained by requiring the theoretical EWs match with the observed values. The calculations have taken account of several broadening mechanisms: natural, thermal, van der Waals damping, and microturbulence. The final abundances, relative to solar abundances derived from the Moon spectrum, are given in Table \ref{tbl:abundance}.

Two types of error sources are taken into account to estimate the uncertainties of the abundances. One is the errors owing to internal uncertainties, in determining EWs and atomic data (mainly $gf$). Here, we only considered the errors from EWs as errors in $gf$ values nearly cancel out when subtracting the solar abundances from the stellar ones.
The other source is uncertainties in stellar parameters ($T_{\rm eff}$, $\log g$, [Fe/H], and $\xi_{t}$). We estimated the impact of the stellar parameters on derived abundances by changing each quantity separately, and leaving the others unchanged. Table \ref{tbl:error1} list the abundance differences when changing the effective temperatures by 100 K, the surface gravities by 0.1 dex, the iron abundances by 0.1 dex, and the microturbulent velocities by 0.2 $ \,\mathrm{km\,s^{-1}}$. Finally, we took the square roots of the quadratic sum of the errors associated to all factors to calculate the total error, which is given as $\sigma_\mathrm{Total}$.
As seen in Table \ref{tbl:error1}, for most of the chemical elements, the uncertainties are less than 0.11 dex. Some neutron capture elements (Sr, Y, and Ba) are sensitive to variations in temperature and microturbulence, for which uncertainty reaches as large as $0.17$ dex.

\begin{table*} \small
\caption{Estimated uncertainties in abundance analysis for the program stars.}
\label{tbl:error1}
\centering
\scalebox{0.6}{
\begin{tabular}{|l|rrrrrr|rrrrrr|rrrrrr|}
\hline\hline
\multicolumn{7}{c}{ BD+68$^\circ$1027}  &  \multicolumn{6}{c}{BD+80$^\circ$670} &  \multicolumn{6}{c}{RE~J0702+129}\\
 \cmidrule(r){2-7} \cmidrule(r){8-13} \cmidrule(r){14-19}
$\Delta [\mathrm{X/H}]$  &
$\frac{\sigma_{EW}}{\sqrt{N}}$ &
$\Delta T_\mathrm{eff}$  &
$\Delta\log g$           &
$\Delta\mathrm{[Fe/H]}$  &
$\Delta\xi_\mathrm{t}$   &
$\sigma_\mathrm{Total}$  &

$\frac{\sigma_{EW}}{\sqrt{N}}$ &
$\Delta T_\mathrm{eff}$  &
$\Delta\log g$           &
$\Delta\mathrm{[Fe/H]}$  &
$\Delta\xi_\mathrm{t}$   &
$\sigma_\mathrm{Total}$   &

$\frac{\sigma_{EW}}{\sqrt{N}}$ &
$\Delta T_\mathrm{eff}$  &
$\Delta\log g$           &
$\Delta\mathrm{[Fe/H]}$  &
$\Delta\xi_\mathrm{t}$   &
$\sigma_\mathrm{Total}$  \\
 &
 &
\small ($+100\mathrm{K}$)    &
 ($+0.1$) &
 ($+0.1$) &
($+0.2$)    &
&
&
\small ($+100\mathrm{K}$)    &
 ($+0.1$) &
 ($+0.1$) &
($+0.2$)    &

    &
    &
\small ($+100\mathrm{K}$)    &
 ($+0.1$) &
 ($+0.1$) &
($+0.2$)

 \\
\hline

$\Delta\mathrm{[	C  	]}$ &	0.03	&	$-$0.05 	&	0.03 	&	$-$0.01 	&	0.00 	&	0.07   	& 0.03 	&	$-$0.05 	&	0.03 	&	0.00 	&	0.00 	&	0.07 	   &   0.03 	&	$-$0.06 	&	0.03 	&	$-$0.01 	&	0.00 	&	0.07 	 \\
$\Delta\mathrm{[	O  	]}$ &	0.06	&	$-$0.07 	&	0.02 	&	$-$0.01 	&	$-$0.01 	&	0.10  & 0.05	&	$-$0.08 	&	0.03 	&	$-$0.01 	&	$-$0.02 	&	0.10 &   0.02	&	$-$0.02 	&	0.11 	&	0.08 	&	0.06 	&	0.15   	   \\
$\Delta\mathrm{[	Na	]}$ &	0.07	&	0.05 	&	$-$0.01 	&	0.00 	&	$-$0.01 	&	0.09 	  & 0.04 	&	0.06 	&	$-$0.01 	&	0.00 	&	$-$0.02 	&	0.08 	 &   0.02 	&	0.06 	&	$-$0.01 	&	0.00 	&	$-$0.02 	&	0.07 	 \\
$\Delta\mathrm{[	Mg	]}$ &	0.02	&	0.05 	&	$-$0.01 	&	0.00 	&	$-$0.01 	&	0.06 		& 0.02 	&	0.04 	&	$-$0.03 	&	0.00 	&	$-$0.01 	&	0.05 	 &   0.03 	&	0.04 	&	$-$0.02 	&	0.01 	&	$-$0.01 	&	0.06 	 \\
$\Delta\mathrm{[	Al	]}$ &	0.04	&	0.04 	&	$-$0.01 	&	0.00 	&	0.00 	&	0.06 			& 0.03 	&	0.04 	&	$-$0.02 	&	0.00 	&	$-$0.01 	&	0.05 	 &   0.03 	&	0.04 	&	$-$0.01 	&	0.00 	&	$-$0.01 	&	0.05 	 \\
$\Delta\mathrm{[	Si	]}$ &	0.01	&	0.03 	&	$-$0.01 	&	0.00 	&	$-$0.01 	&	0.03 	  & 0.02 	&	0.02 	&	0.00 	&	0.01 	&	$-$0.02 	&	0.04 	   &   0.03 	&	0.01 	&	0.00 	&	0.01 	&	$-$0.02 	&	0.04 	   \\
$\Delta\mathrm{[	K  	]}$ &	0.03	&	0.09 	&	$-$0.04 	&	0.01 	&	$-$0.03 	&	0.11 	  & 0.03 	&	0.09 	&	$-$0.05 	&	0.01 	&	$-$0.04 	&	0.11 	 &   0.03 	&	0.10 	&	$-$0.05 	&	0.03 	&	$-$0.03 	&	0.12   \\
$\Delta\mathrm{[	Ca	]}$ &	0.02	&	0.07 	&	$-$0.02 	&	0.00 	&	$-$0.02 	&	0.08 	  & 0.02 	&	0.07 	&	$-$0.02 	&	0.01 	&	$-$0.04 	&	0.09 	 &   0.05 	&	0.07 	&	$-$0.01 	&	0.00 	&	$-$0.03 	&	0.09   \\
$\Delta\mathrm{[	ScII	]}$ &	0.02	&	0.01 	&	0.03 	&	0.02 	&	$-$0.03 	&	0.05 	  & 0.02 	&	0.00 	&	0.04 	&	0.03 	&	$-$0.03 	&	0.06 	   &   0.03 	&	0.00 	&	0.04 	&	0.03 	&	$-$0.04 	&	0.07     \\
$\Delta\mathrm{[	TiI	]}$ &	0.02	&	0.10 	&	$-$0.01 	&	0.00 	&	$-$0.03 	&	0.11 	  & 0.02 	&	0.10 	&	$-$0.01 	&	0.00 	&	$-$0.04 	&	0.11 	 &   0.04 	&	0.12 	&	$-$0.01 	&	0.00 	&	$-$0.05 	&	0.14   \\
$\Delta\mathrm{[	TiII	]}$ &	0.03	&	0.01 	&	0.03 	&	0.02 	&	$-$0.03 	&	0.06 	  & 0.02 	&	0.00 	&	0.04 	&	0.03 	&	$-$0.04 	&	0.07 	   &   0.04 	&	0.00 	&	0.04 	&	0.03 	&	$-$0.03 	&	0.07     \\
$\Delta\mathrm{[	V  	]}$ &	0.03	&	0.10 	&	0.00 	&	0.00 	&	$-$0.01 	&	0.10 			& 0.04 	&	0.10 	&	0.00 	&	0.00 	&	$-$0.02 	&	0.11 	   &   0.02 	&	0.11 	&	0.00 	&	0.00 	&	$-$0.02 	&	0.11 	   \\
$\Delta\mathrm{[	CrI	]}$ &	0.04	&	0.08 	&	$-$0.02 	&	0.00 	&	$-$0.03 	&	0.10 		& 0.03 	&	0.09 	&	$-$0.02 	&	0.01 	&	$-$0.04 	&	0.11 	 &   0.03 	&	0.07 	&	$-$0.01 	&	0.00 	&	-0.03 	&	0.08 	 \\
$\Delta\mathrm{[	CrII	]}$ &	0.01	&	$-$0.01 	&	0.03 	&	0.01 	&	$-$0.02 	&	0.04  & 0.02 	&	$-$0.02 	&	0.03 	&	0.02 	&	$-$0.04 	&	0.06 	 &   0.01 	&	$-$0.03 	&	0.03 	&	0.02 	&	$-$0.05 	&	0.07 	 \\
$\Delta\mathrm{[	Mn	]}$ &	0.02	&	0.08 	&	$-$0.02 	&	0.00 	&	$-$0.02 	&	0.09 		& 0.05 	&	0.08 	&	$-$0.02 	&	0.01 	&	$-$0.05 	&	0.11 	 &   0.03 	&	0.07 	&	0.00 	&	0.00 	&	$-$0.02 	&	0.08 	   \\
$\Delta\mathrm{[	FeI	]}$ &	0.01	&	0.08 	&	$-$0.01 	&	0.00 	&	$-$0.04 	&	0.09 		& 0.01 	&	0.07 	&	$-$0.01 	&	0.00 	&	$-$0.05 	&	0.09 	 &   0.01 	&	0.07 	&	0.00 	&	0.01 	&	$-$0.04 	&	0.08 	   \\
$\Delta\mathrm{[	FeII	]}$ &	0.01	&	$-$0.01 	&	0.03 	&	0.02 	&	$-$0.03 	&	0.05  & 0.01 	&	$-$0.02 	&	0.03 	&	0.03 	&	$-$0.05 	&	0.07 	 &   0.03 	&	$-$0.04 	&	0.04 	&	0.03 	&	$-$0.03 	&	0.08 	 \\
$\Delta\mathrm{[	Co	]}$ &	0.03	&	0.09 	&	0.00 	&	0.02 	&	0.00 	&	0.10 				&  0.02   &  0.06  &    0.00   &   0.00  &   $-$0.02    &   0.07   &          0.03 	&	0.09 	&	0.01 	&	0.01 	&	$-$0.01 	&	0.10 	   \\
$\Delta\mathrm{[	Ni	]}$ &	0.02	&	0.07 	&	$-$0.01 	&	0.00 	&	$-$0.02 	&	0.08 		& 0.02 	&	0.07 	&	$-$0.01 	&	0.01 	&	$-$0.03 	&	0.08   &   0.02 	&	0.06 	&	0.00 	&	0.01 	&	$-$0.03 	&	0.07 	   \\
$\Delta\mathrm{[	Cu	]}$ &	0.02	&	0.07 	&	$-$0.01 	&	0.00 	&	$-$0.03 	&	0.08 		& 0.03 	&	0.09 	&	$-$0.01 	&	0.02 	&	$-$0.08 	&	0.13   &   0.03 	&	0.03 	&	0.01 	&	0.05 	&	$-$0.07 	&	0.10 	   \\
$\Delta\mathrm{[	Sr	]}$ &	0.01	&	0.10 	&	$-$0.02 	&	-0.01 	&	$-$0.08 	&	0.13  & 0.02 	&	0.11 	&	$-$0.03 	&	0.01 	&	$-$0.10 	&	0.15   &   0.03 	&	0.10 	&	$-$0.01 	&	0.02 	&	$-$0.09 	&	0.14 	 \\
$\Delta\mathrm{[	Y  	]}$ &	0.02	&	0.02 	&	0.03 	&	0.03 	&	-0.06 	&	0.08 			& 0.01 	&	0.03 	&	0.03 	&	0.04 	&	$-$0.06 	&	0.08 	   &   0.05 	&	0.12 	&	$-$0.02 	&	0.01 	&	$-$0.11 	&	0.17 	 \\
$\Delta\mathrm{[	Zr	]}$ &	0.03	&	0.01 	&	0.04 	&	0.02 	&	$-$0.01 	&	0.06 			& 0.03 	&	0.00 	&	0.04 	&	0.03 	&	$-$0.02 	&	0.06 	   &   0.05 	&	0.01 	&	0.03 	&	0.04 	&	$-$0.04 	&	0.08 	   \\
$\Delta\mathrm{[	Ba	]}$ &	0.03	&	0.05 	&	$-$0.01 	&	0.04 	&	$-$0.05 	&	0.09 		& 0.02 	&	0.03 	&	0.01 	&	0.04 	&	-0.08 	&	0.10 	   &   0.02 	&	0.13 	&	0.00 	&	0.00 	&	$-$0.01 	&	0.13 	   \\
$\Delta\mathrm{[	La	]}$ &	0.04	&	0.04 	&	0.04 	&	0.03 	&	$-$0.02 	&	0.08 			&  -     &   -  &   -   & -     &    -    &   -       &  0.03 	&	0.03 	&	0.04 	&	0.03 	&	0.00 	&	0.07 	     \\
$\Delta\mathrm{[	Ce	]}$ &	0.03	&	0.04 	&	0.03 	&	0.02 	&	$-$0.05 	&	0.08 		  & 0.03 	&	0.02 	&	0.04 	&	0.03 	&	$-$0.03 	&	0.07     &   0.05 	&	0.02 	&	0.04 	&	0.03 	&	$-$0.02 	&	0.08 	   \\
$\Delta\mathrm{[	Nd	]}$ &	0.01	&	0.04 	&	0.04 	&	0.03 	&	$-$0.02 	&	0.07 			&	- 	&	- 	&	- 	&	- 	&	- 	&	-     &   0.01 	&	0.03 	&	0.04 	&	0.04 	&	$-$0.01 	&	0.07 	   \\

\hline
\end{tabular}
}
\end{table*}

 \section{Notes on individual stars}

 \subsection{BD+68$^\circ$1027}
 BD+68$^\circ$1027 is a high proper-motion star based on $SIMBAD$. \cite{1999Prieto} estimated \mbox{its} $T_{\rm eff}$, $\log g$ and mass are 5754 K, 4.57 and 0.93 $M_\odot $, respectively. Stellar parameters for this star from Table \ref{para} are compared with these values, and are in good agreement.
Among three Ba dwarfs in this work, it is the most metal-poor and abundant in s-process elements.
Li line at 6707.76 \AA\ is almost absent in this star, and the measured EW is $\sim$1.2 m\AA.

 \begin{figure*}
\begin{center}

    \includegraphics[width=14cm]{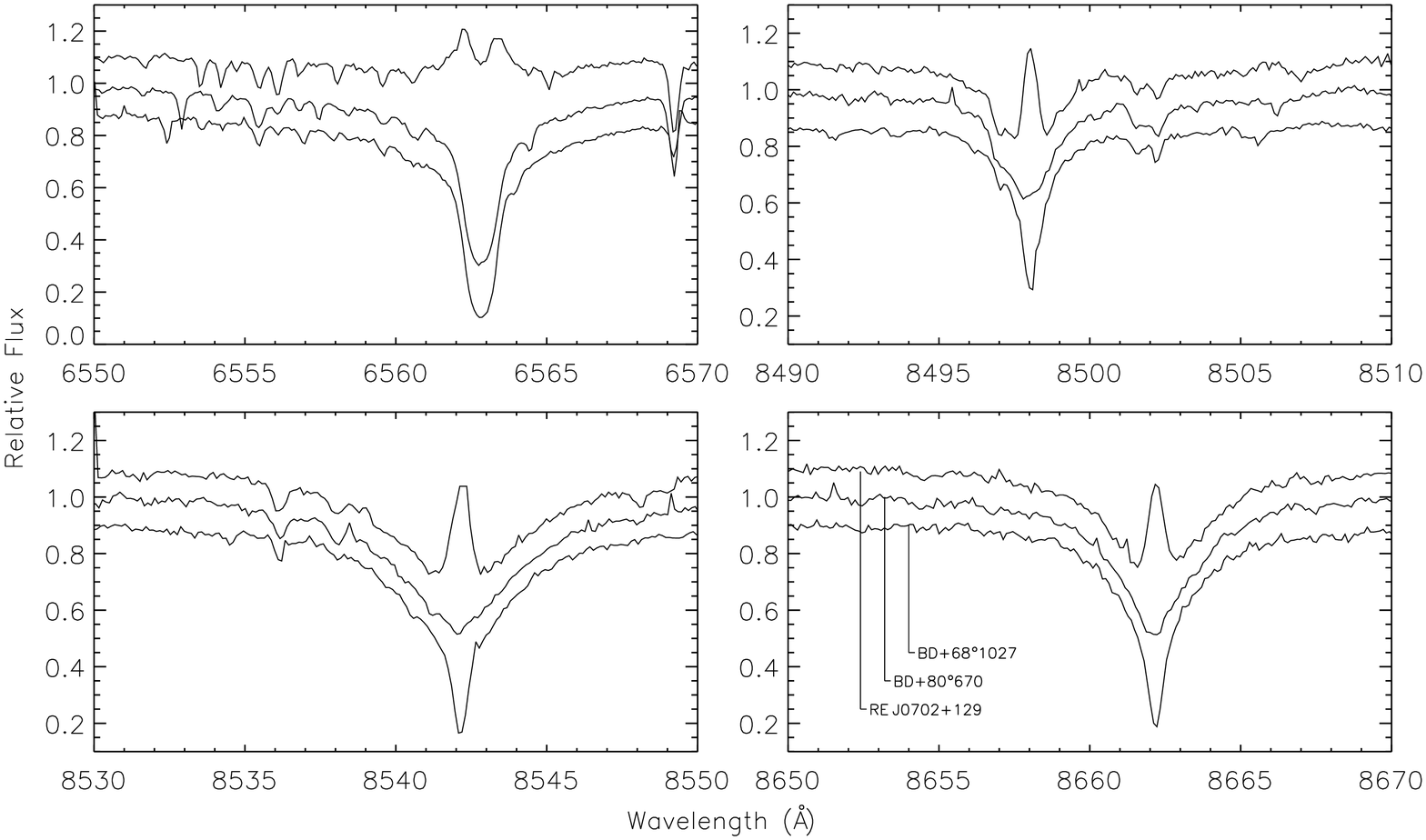}
    \caption{
    Spectra of sample stars are plotted in the region of H$\alpha$ and Ca~{\sc ii} IR triplet. Note strong emission features are evident in RE~J0702+129.}

\label{Halpha}
\end{center}
\end{figure*}

\subsection{BD+80$^\circ$670}
BD+80$^\circ$670 is also a high proper-motion star based on $SIMBAD$. \cite{2009Guillout} determined its atmospheric parameters, $T_\mathrm{eff}(Spec)$ = 5745, $\log g(Spec)$ = 4.30, [Fe/H] = 0.06, and our values are in good agreement with them (see in Table \ref{para}).
None of the lines of La and Nd are clean enough to derive their abundances as part of s-process elements.
The Li\,{\sc i} around 6707.76 {\AA} was very weak with measured EW of $\sim$3.00 m\AA, consistent with \cite{2009Guillout} determination of EW of 6707.76 around 3.50 m\AA.

\subsection{RE~J0702+129}
This star is listed as an X-ray binary in SIMBAD, due to the detection of strong X-ray flux with hardness ratio of $-$0.73 \citep{1997Vennes}.  According to \cite{1997Vennes}, it hosts a WD companion and hence falls in the category of SLSs, catalogued by \cite{Holberg2013}. Further, \cite{2010Barstow} found this star is in triple system.  There are no previous determination of atmospheric parameters and abundances for this star. Abundance analysis from this work shows that C, O, Na and K are relatively higher than solar values.
\cite{1986Smith} and \cite{1993Lambert} concluded that Ba giant and \mbox{ dwarf stars} are Li-poor, and the absorption feature at 6707 \AA\ might
be dominated by Ce line \citep{2002Reyniers}. Interestingly, for this star, we derived its Li abundance log $\epsilon$ (Li) $\sim$ 1.55 (on a scale of log $\epsilon$ (H) = 12) using spectrum synthesis method. For the other two sample stars in this work, the lithium lines around 6707 \AA\ were very weak, which are consistent with the above conclusion.

The spectrum of this star shows strong emission in H$\alpha$ and Ca~{\sc ii} IR triplet (see Figure~\ref{Halpha}), indicating the presence of high level chromospheric activity. Such activity was previously noticed by \citet{1997Vennes} based on emission feature at the core of Ca H \& K and H$\alpha$.
 In fact, this star was identified as rotationally variable based on variation in its light curve due to starspots \citep{2012Kiraga}.
 The rotational period, $P_{\rm rot}$ = 2.81 days, suggests this star is fast rotator and corresponds to a Rossby number Ro $\approx$ 0.18 (Ro = $P_{\rm rot}$/$\tau$),
 where the convective turnover time $\tau\sim16$ days was estimated from its mass using the correlation between
convective turnover time and mass given by \cite{2011Wright}.
Its Rossby number is close to that of saturation (e.g., Ro$_{sat}$ = 0.13, \cite{2011Wright}), also indicates high activity level as shown by chromospheric emission lines. This star also show coronal emission $L_{X}/L_{bol}$ = $-$2.74 \citep{2012Kiraga},
a value even larger than the saturation level of coronal activity $L_{X}/L_{bol}\sim -3$ \citep{2011Wright,2014Reiners}.
 Compared with the spectrum of \citet{1997Vennes}, it appears that H$\alpha$ emission line is slightly stronger (e.g., shallower absorption in the center) in our spectrum, suggesting this star may be more active than before during our observing run, e.g., a flare-like event, which may be partly responsible for strong Li line similar to 2RE J0743+224, a chromospherically active binary in
which Li enhancement was detected during a long-duration flare \citep{1998Montes}. The activity levels in this star suggests it is an interesting object to consider for continuous monitoring to check the variations in H$\alpha$ and spectral lines of other elements.

\section{Discussion}

 \subsection{Ba Classification}
\begin{figure}
\begin{center}

   \includegraphics[width=\columnwidth]{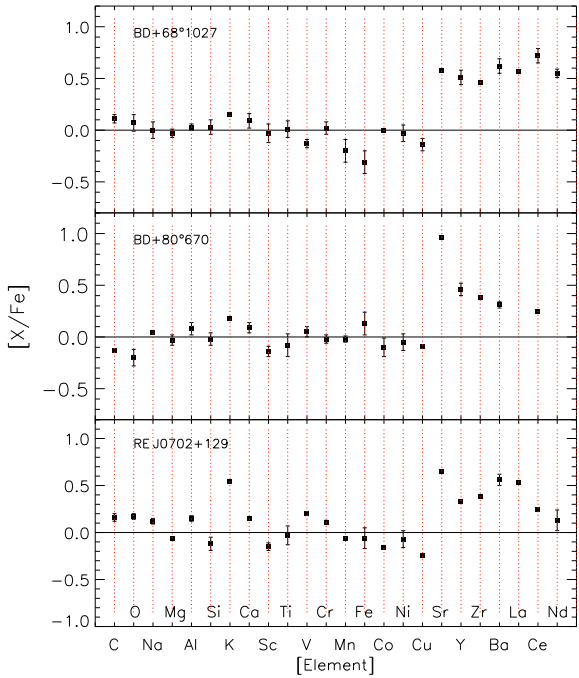}
    \caption{
    Element to iron ratio of 24 elements are shown along with error bars ($\sigma_\mathrm{Total}$ of Table \ref{tbl:error1}). Note the enhancement in s-process elements are clearly seen in all three stars.
    }

\label{xFe}
\end{center}
\end{figure}

  The abundance patterns of BD+68$^\circ$1027, BD+80$^\circ$670 and RE~J0702+129 are shown in Figure \ref {xFe}. The three stars show different levels of enhancement in s-process elements with [s/Fe] (the mean value of Sr, Y, Zr, Ba, La, Ce and Nd) ratios  of 0.57, 0.47 and 0.41, respectively. As \cite{Castro2016} have pointed out, there is no clear agreement in the literature on how high should be the [s/Fe] ratio for a \mbox{star} to be considered as a barium or even a mild barium \mbox{star}. \cite{Sneden1981}, \cite{Porto1997} and \cite{Rojas2013} discovered some mild barium stars with the minimum value of [s/Fe] ratios, 0.21, 0.23 and 0.34, respectively, and \cite{Castro2016} assumed a value of +0.25 as a minimum [s/Fe] ratio for a star to be considered as a barium star. Considering the above criteria, we classified BD+68$^\circ$1027, RE~J0702+129 and BD+80$^\circ$670 as mild Ba dwarf stars.

\subsection{Comparison with known Ba dwarfs and CH subgiants}
\begin{figure}
\begin{center}

	\includegraphics[width=\columnwidth]{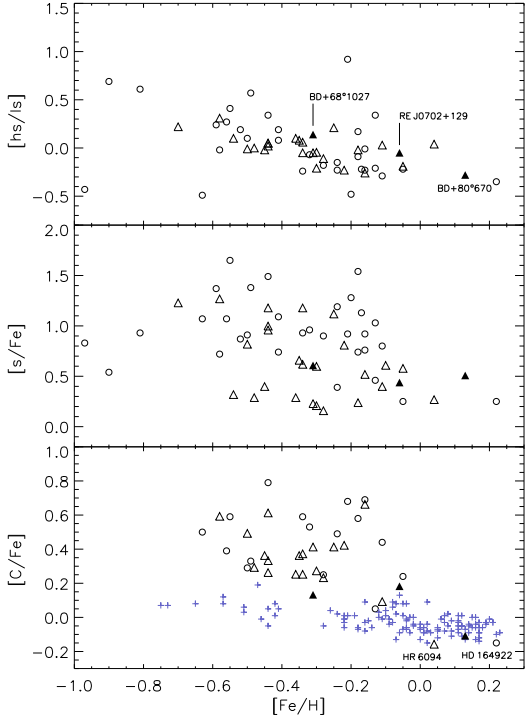}
    \caption[Cap for listoffigures]{
    The ratios of [hs/ls], [s/Fe] and [C/Fe] are plotted against metallicity. Open triangles and circles: Ba dwarfs and CH subgiants from literature (see Table \ref{tab:infor}). Filled triangles: Ba dwarfs analyzed in this work. Plus symbols: field FGK dwarfs from \cite{Silva2015}.
    }
\label{sFe}
\end{center}
\end{figure}

With the use of existing abundances of Ba dwarfs, \cite{2005Pereira} suggested both CH subgiants and Ba \mbox{dwarfs} show similar abundance patterns and may share same physical mechanism for their s-process elements \mbox{enrichment}.
\cite{2017Escorza} located Ba and CH stars on Hertzsprung-Russell diagram based on parallaxes provided by the Tycho-Gaia Astrometric Solution and found no clear distinction between Ba dwarfs and CH subgiants with surface gravities ($\log g$) greater than 4.0 dex. So, we collected the Ba dwarfs and the CH subgiants whose $\log g$ are greater than 3.5 with available abundances of s-process elements from various literature (see Table \ref{tab:infor}), and compared the [s/Fe], [hs/ls] and [C/Fe] ratios of our sample with them.
The [hs/ls] ratio has been widely used to indicate the s-process efficiency since \cite{1991LuckL}. [hs] and [ls] stand for the mean abundance of `heavy' s-process elements (Ba, La, Ce and Nd) at the Ba peak and the same for the `light' s-process elements (Sr, Y and Zr) at the Zr peak, respectively.
In Figure~\ref{sFe}, [hs/ls] ratio shows an anticorrelation with metallicity, and the data of our three Ba dwarfs fit well in this diagram. For RE~J0702+129 and BD+80$^\circ$670, the small values of their [hs/ls] support that the neutron capture process becomes less efficient at solar metallicity.
Also, [s/Fe] ratio increases with decrease in metallicity, despite the large dispersion (See Figure~\ref{sFe}).

\begin{figure}
\begin{center}

	\includegraphics[width=\columnwidth]{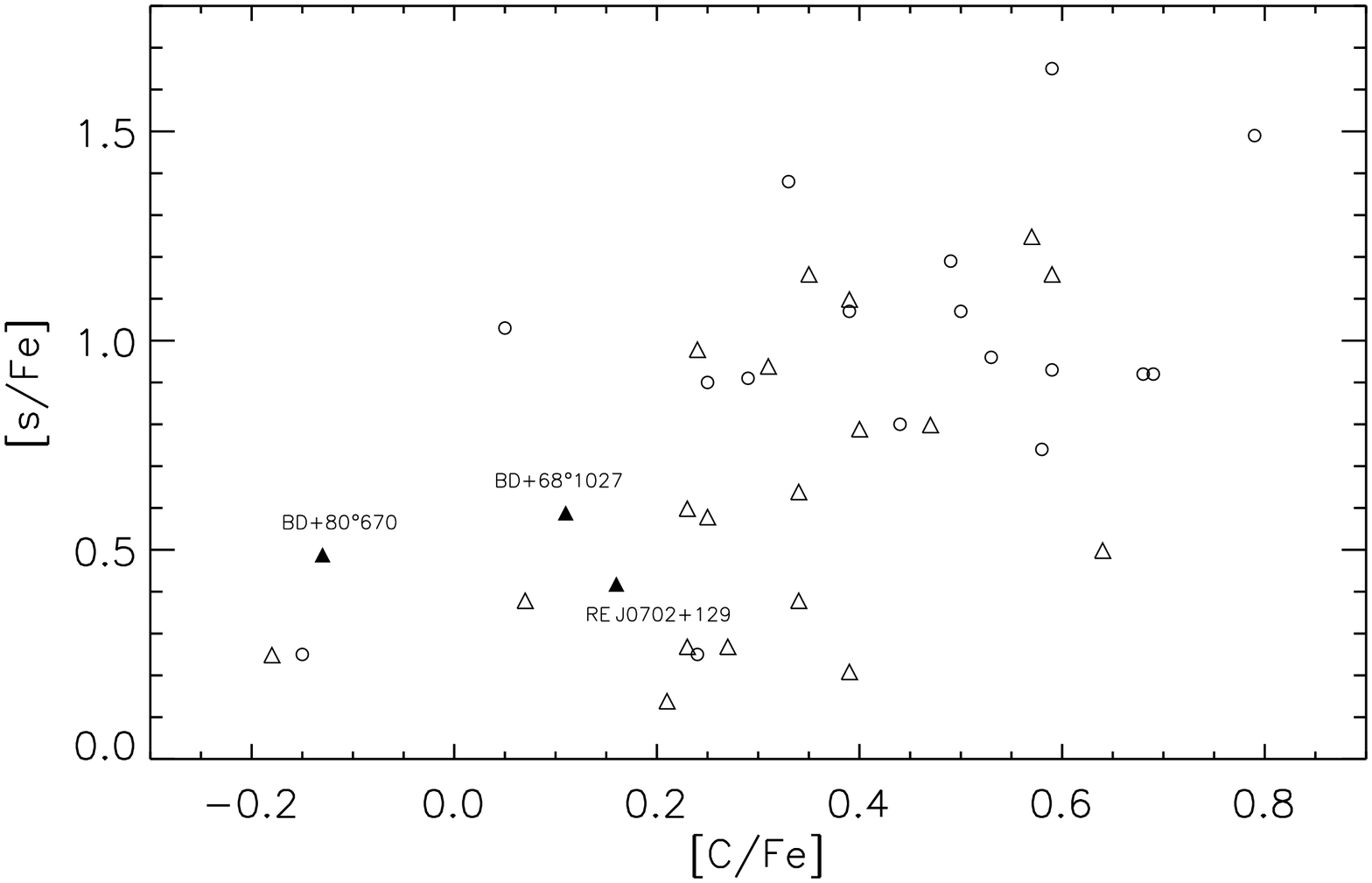}
    \caption{
    The ratio of [s/Fe] is plotted against carbon abundance. Symbols have the same meaning as in Figure \ref{sFe}.
    }
\label{sC}
\end{center}
\end{figure}

The Ba stars were characterised with C \mbox{enrichment} when they were first identified by \cite{Bidelman1951}, and in the classical Ba star scenario, the TP-AGB star enriches its envelope with carbon that was produced in the form of $^1$$^2$C by shell He burning, in addition to the neutron capture elements \citep{1997Porto}. Later studies show that some Ba stars possess weak CH, C$_{\rm 2}$ and CN molecular band strengths \citep{Sneden1981}.
From Figure~\ref{sFe}, we can see that BD+68$^\circ$1027 and RE~J0702+129 show slight overabundance of carbon ([C/Fe] $\sim$ 0.11 and 0.16, respectively) compared with disk stars of similar metallicity.
BD+80$^\circ$670 is showing deficiency in C similar to other two metal-rich Ba dwarfs, HR~6094 and HD~164922, but still follows the trend of normal field \mbox{dwarfs}. \cite{1997Porto} suggested C deficiency in HR~6094 might be due to companion AGB star attributed to hot bottom burning episode. The WD companion mass of BD+80$^\circ$670 is 0.81 $M_\odot$, which is close to what one would expect for hot bottom burning taking place in AGB stars with core masses $M_\mathrm{c} \geq 0.85 M_\odot$ \citep{1992Boothroyd}.
From Figure~\ref{sFe}, it is clear that [C/Fe] shows anticorrelation with metallicity, and carbon abundance show a positive correlation with s-process elements, as shown in Figure~\ref{sC}.

\cite{Vitense2000} and \cite{2000North} indicated that Ba dwarfs will become Ba giants after they \mbox{evolve} off main sequence. However, there are some open questions about this hypothesis.
For example, \cite{Castro2016} and \cite{2017Escorza} showed the metallicity distribution of Ba giants in which quite a few samples occupied metal-rich side $-0.1 < [Fe/H] < 0.1$. From Figure~\ref{sFe}, it is clear that the number of Ba dwarfs in the same metallicity range is very small. If the Ba giants evolved from Ba dwarfs, then more Ba dwarfs at near solar metallicity need to be identified.
RE~J0702+129 and BD+80$^\circ$670 are increasing the Ba dwarfs number \mbox{towards} metal-rich range and are of positive significance for the above hypothesis.
The mass of known Ba dwarfs are found to be slightly below 1 $M_\odot $ to above 2 $M_\odot $ (e.g.\citealt{2017Escorza} ). The estimated masses of our target stars (0.93$M_{\odot}$, 0.93$M_{\odot}$ and 1.05$M_{\odot}$, respectively) are within this range, which are still lower than Ba giants. However, the deficit of high-mass Ba dwarfs is probably caused by an observational selection \mbox{effect}. Massive Ba dwarfs are probably concealed among Am-Fm stars and difficult to identify \citep{2000North}.

\subsection{Comparison with AGB models}

 \cite{Hurley2000} predicted that the minimum WD mass for its progenitor reach the AGB phase is at least 0.51 $M_\odot $, i.e, if WD mass is less than the threshold value, the s-process synthesis didn't occur in its progenitor and couldn't transfer s-process enrichment materials to its companion.
  For our three target stars, except BD+68$^\circ$1027 which has no WD mass information, the masses of WD companions for BD+80$^\circ$670 and RE~J0702+129 are 0.81 and 0.57 $M_\odot $, respectively, which are consistent with the prediction of \cite{Hurley2000}. The progenitor masses (during AGB) estimated using the formulas given by \cite{2008Catal} for the WD companions of the two target stars are 3.59 and 1.47 $M_\odot $, respectively, which confirms the predicted range of WDs arising from AGB progenitors of mass (1.5 $\sim$ 4 $M_{\odot}$) \citep{1999Dominguez}.

The abundance patterns of s-process elements in these stars are compared with solar metallicity models from \cite{2016Karakas}, wherein they provided surface abundances ([X/Fe]) of s-process elements predicted at every thermal pulse (TP) in low mass AGB stars (See Figure~8). As barium stars owe their s-process \mbox{enhancement} to mass transfer, their envelopes were mixed and contaminated by the s-rich materials from the AGB donors. The masses of BD+80$^\circ$670 (1.05 $M_{\odot}$) and RE~J0702+129 (0.93 $M_{\odot}$) suggest their envelopes are convective. Based on MESA (Modules for Experiments in Stellar Astrophysics) \mbox{models} \citep{2011Paxton, 2013Paxton, 2015Paxton}, we have estimated the masses of the outer convective zones for BD+80$^\circ$670 and RE~J0702+129 to be 0.0451 $M_{\odot}$ and 0.0076 $M_{\odot}$, respectively.

Let us define the pollution factor $r$ as $\mathrm{M}_{acc}$/$\mathrm{M}_{env}$, where $\mathrm{M}_{env}$ is the mass of the envelope in a Ba dwarf after pollution, and $\mathrm{M}_{acc}$ is the accreted mass from its companion AGB star. The pollution factor $r$ is then chosen in order to match the [X/Fe] of the observed spectroscopic data. The theoretical s-process abundances with corresponding pollution factors are calculated and their associated \mbox{parameters} are given in Table~A1. (see Appendix for details). Note that the level of s-process enrichment in the accreted materials changes continuously due to various thermal pulses in the AGB star, and we use a parameter $\gamma$ to describe this `step process' \citep{Boffin1988}. The calculation of related parameters and the dilution process are described in details in the Appendix. In Figure~8, r=1.00 means the theoretical abundances in the \mbox{AGB} star (the [X/Fe]$^{fnl}$ in Table~A1), and we list it for reference.
The AGB models with initial masses of 1.5 \& 3.5 $M_{\odot}$ are chosen for RE~J0702+129 and BD+80$^\circ$670, respectively, that are close to their companion AGB masses (1.47 \& 3.59 $M_{\odot}$).
The adopted pollution factors for model s-process abundances that are matching observing abundances of BD+80$^\circ$670 (between 0.06 and 0.11) are within predicted range (0.05$\le r \le $0.2) for mild Ba stars\citep{Han1995}. However, observed [Sr/Fe] is higher and required r of 0.7 which is in range of strong Ba stars. For RE~J0702+129 the observed abundances are matching closely with pollution factor of 0.4, a higher value than predicted range for mild Ba stars, suggests s-process efficiency is less in low mass AGB stars and requires more pollution. However, it is noticed that [Sr/Fe] is higher than adopted model values.
 The accretion efficiencies for both the stars in Table~A2 show they are within predicted range of \cite{2017arXivliu}, wherein suggested that accretion efficiency onto the secondary star varies from about 0.1$\%$ to 8$\%$ for mass ratios between 0.05 and 1.0.

 \begin{figure}
\begin{center}
	\includegraphics[width=\columnwidth]{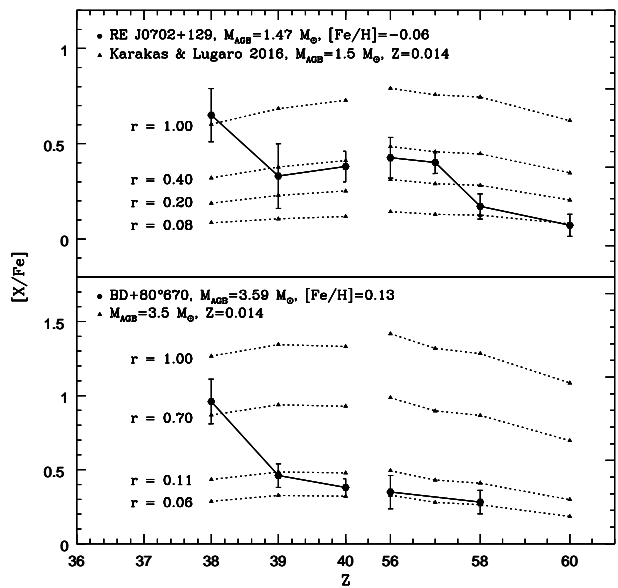}
    \caption{[X/Fe] vs. Z for RE~J0702+129 and BD+80$^\circ$670, compared to theoretical surface abundances predicted for AGB stars of Karakas \& Lugaro (2016). For the two sample stars, ${M_{AGB}}$ means their estimated AGB companion mass. Dotted lines represent theoretical surface abundances of Ba stars with different pollution factor r.}
\label{fig8}
\end{center}
\end{figure}

\section{Conclusions}
 Chemical composition analysis of three dwarf primaries \mbox{(BD+68$^\circ$1027, BD+80$^\circ$670 and RE~J0702+129)} of SLSs show enhancement in s-process elements, and based on their [s/Fe] ratios (0.57, 0.47, and 0.41) we categorised them as mild Ba stars. RE~J0702+129 and BD+80$^\circ$670 increase the number of Ba dwarfs with near solar-metallicity that indicates Ba \mbox{dwarfs} span large metallicity range similar to Ba giants. The estimated masses (0.93, 1.05 and 0.93 ${M_\odot}$) for these three target stars are within the range noticed in Ba dwarfs (\mbox{slightly} below 1 ${M_\odot}$ to above 2 ${M_\odot}$). The ratios of [hs/ls], [s/Fe] and [C/Fe] of our sample fit well with the observed trend of Ba dwarfs and it is clear that they show anticorrelation with metallicity.

Based on WD companion masses (0.57 and 0.81 ${M_\odot}$) of RE~J0702+129 and BD+80$^\circ$670, their progenitor \mbox{AGB} masses are estimated as 1.47 and 3.59 ${M_\odot}$, respectively, which perfectly fit to the predicted range of expected progenitor AGB masses (1.5$\sim$4 ${M_\odot}$) of unseen WDs around Ba dwarfs.
 The abundance profiles of the s-process elements from RE~J0702+129 and BD+80$^\circ$670 are compared with models of corresponding companion AGB masses from \cite{2016Karakas}, and found the estimated accretion efficiencies and pollution factors are consistent with the prediction of \cite{2017arXivliu} and \cite{Han1995}.

In contrast, our study confirms the existence of WD companions to Ba dwarfs and supports the origin of s-process enrichment in Ba dwarfs via McClure hypothesis similar to Ba giants, in which host WD companion during its TP-AGB phase contaminates primary star atmosphere with s-process elements through mass transfer.

It is important to understand the role of companion masses, the orbital period of the binary system and the metallicity in observed levels of s-process enrichment in Ba stars, and we discuss these relations in subsequent publication based on the behaviour of various elements in 21 dwarf and giant primaries (Ba and normal) of SLSs.

\begin{table*}

\centering
\caption{Atmospheric parameters, stellar mass, Carbon, and s-process abundances of Ba dwarfs and CH subgiants.    }
\label{tab:infor}
\scalebox{0.8}{
\begin{tabular}{l|ccrcrrrrrrrrrlc}
\hline\hline

Star &
$M_\mathrm{p}/{M_\odot}$    &
$T_\mathrm{eff}$(K)  &
$\mathrm{[Fe/H]}$  &
$\log g$    &
$\mathrm{[C/Fe]}$   &
$\mathrm{[Sr/Fe]}$   &
$\mathrm{[Y/Fe]}$  &
$\mathrm{[Zr/Fe]}$   &
$\mathrm{[Ba/Fe]}$  &
$\mathrm{[La/Fe]}$   &
$\mathrm{[Ce/Fe]}$   &
$\mathrm{[Nd/Fe]}$  &
$\mathrm{[s/Fe]}$  &
remark  &
ref  \\

\hline

 BD+68$^\circ$1027	&	0.93 	&	5919	&	$-$0.31	&	4.48	&	0.11	&	0.58	&	0.51	&	0.46	&	0.62	&	0.57	&	0.72	&	0.55	&	0.57 	&	dwarf	&	this work	\\
{\bf BD+80$^\circ$670}	&	1.05 	&	5880	&	{\bf 0.13} 	&	4.33	&	{\bf $-$0.13}	&	0.96	&	0.46	&	0.38	&	0.31	&	-	&	0.25	&	-	&	{\bf 0.47} 	&	dwarf	&	this work	\\
RE~J0702+129	&	0.93	&	5531	&	$-$0.06 	&	4.20	&	0.16	&	0.65 	&	0.33 	&	0.38 	&	0.56 	&	0.53 	&	0.25 	&	0.13 	&	0.41 	&	dwarf	&	this work	\\
\hline
HR~107	&	1.2	&	6440	&	$-$0.34 	&	4.10 	&	0.23	&	0.53 	&	0.60 	&	0.59 	&	0.95 	&	0.67 	&	0.41 	&	0.32 	&	0.58 	&	dwarf	&	1	\\
HD~8270   	&	0.9	&	5940	&	$-$0.44 	&	4.20 	&	0.31	&	0.87 	&	0.95 	&	0.92 	&	1.11 	&	1.00 	&	0.83 	&	0.73 	&	0.92 	&	dwarf	&	1	\\
HD~13551  	&	0.9	&	5870	&	$-$0.44 	&	4.00 	&	0.24	&	0.88 	&	1.08 	&	0.96 	&	1.16 	&	0.99 	&	0.91 	&	0.73 	&	0.96 	&	dwarf	&	1	\\
HD~76225  	&	1.4	&	6110	&	$-$0.34 	&	3.80 	&	0.35	&	1.17 	&	1.17 	&	1.22 	&	1.35 	&	1.22 	&	1.06 	&	0.77 	&	1.14 	&	dwarf	&	1	\\
HD~87080  	&	1.2	&	5460	&	$-$0.49 	&	3.70 	&	0.33	&	0.99 	&	1.11 	&	1.08 	&	1.48 	&	1.74 	&	1.73 	&	1.56 	&	1.38 	&	sgCH	&	1	\\
HD~89948  	&	1.0	&	6010	&	$-$0.28 	&	4.30 	&	0.25	&	1.03 	&	1.02 	&	0.94 	&	0.99 	&	0.93 	&	0.71 	&	0.65 	&	0.90 	&	sgCH	&	1	\\
HD~106191 	&	1.0	&	5890	&	$-$0.22 	&	4.20 	&	0.40	&	0.80 	&	0.91 	&	1.07 	&	0.88 	&	0.66 	&	0.61 	&	0.48 	&	0.77 	&	dwarf	&	1	\\
HD~107574 	&	1.4	&	6400	&	$-$0.56 	&	3.60 	&	0.39	&	0.85 	&	0.96 	&	0.95 	&	1.71 	&	1.16 	&	1.02 	&	0.86 	&	1.07 	&	sgCH	&	1	\\
HD~123585 	&	1.1	&	6350	&	$-$0.44 	&	4.20 	&	0.79	&	1.16 	&	1.34 	&	1.41 	&	1.79 	&	1.65 	&	1.69 	&	1.41 	&	1.49 	&	sgCH	&	1	\\
HD~150862 	&	1.1	&	6310	&	$-$0.11 	&	4.60 	&	0.44	&	0.76 	&	1.08 	&	1.07 	&	1.03 	&	0.80 	&	0.55 	&	0.34 	&	0.80 	&	sgCH	&	1	\\
HD~188985 	&	1.1	&	6090	&	$-$0.25 	&	4.30 	&	0.39	&	1.02 	&	1.02 	&	0.92 	&	1.20 	&	1.16 	&	1.23 	&	1.04 	&	1.08 	&	dwarf	&	1	\\
HD~222349 	&	1.2	&	6130	&	$-$0.58 	&	3.90 	&	0.57	&	0.98 	&	1.03 	&	1.23 	&	1.38 	&	1.35 	&	1.40 	&	1.26 	&	1.23 	&	dwarf	&	1	\\
BD+18$^\circ$5215	&	1.1	&	6300	&	$-$0.44 	&	4.20 	&	0.59	&	1.13 	&	1.01 	&	1.28 	&	1.46 	&	1.19 	&	1.11 	&	0.83 	&	1.14 	&	dwarf	&	1	\\
 BD$-$03$^\circ$3668	&	0.8-1.0	&	5220	&	$-$0.55 	&	3.70 	&	0.59	&	-	&	1.30 	&	1.45 	&	1.64 	&	1.79 	&	1.73 	&	1.98 	&	1.65 	&	sgCH	&	2	\\
{\bf HR~6094}	&	-	&	5859	&	{\bf 0.04} 	&	4.63	&	{\bf $-$0.18} 	&	-	&	0.22 	&	0.24 	&	0.37 	&	-	&	0.06 	&	0.27 	&	{\bf 0.23} 	&	dwarf	&	3	\\
HD~26367	&	-	&	6160	&	$-$0.05 	&	4.00 	&	-	&	 0.65 	&	-	&	-	&	0.42 	&	-	&	-	&	-	&	0.54 	&	dwarf	&	4	\\
 HD~22589	&	0.8	&	5600	&	$-$0.16 	&	3.80 	&	0.64	&	-	&	0.72 	&		&	0.75 	&	0.56 	&	0.29 	&	0.07 	&	0.48 	&	dwarf	&	5	\\
HD~50264	&	1.0	&	5800	&	$-$0.34 	&	4.20 	&	0.59	&	-	&	0.88 	&	1.29 	&	1.25 	&	0.92 	&	0.76 	&	0.47 	&	0.93 	&	sgCH	&	6	\\
HD~11377	&	-	&	6000	&	$-$0.05 	&	4.10 	&	0.24	&	-	&	0.29 	&	0.43 	&	0.03 	&	-	&	-	&	0.26 	&	0.25 	&	sgCH	&	7	\\
HD~125079	&	-	&	5300	&	$-$0.16 	&	3.50 	&	0.69	&	-	&	1.22 	&	0.84 	&	0.75 	&	-	&	-	&	0.85 	&	0.92 	&	sgCH	&	7	\\
HD~182274	&	-	&	6000	&	$-$0.18 	&	4.50 	&	0.58	&	-	&	0.77 	&	0.79 	&	0.59 	&	-	&	-	&	0.80 	&	0.74 	&	sgCH	&	7	\\
HD~219116	&	-	&	5300	&	$-$0.32 	&	3.50 	&	0.53	&	-	&	1.12 	&	0.87 	&	0.89 	&	-	&	-	&	0.97 	&	0.96 	&	sgCH	&	7	\\
 BD$-$18$^\circ$255	&	-	&	6800	&	$-$0.18 	&	4.30 	&	-	&	-	&	1.58 	&	1.30 	&	-	&	1.77 	&	1.46 	&	1.59 	&	1.54 	&	sgCH	&	8	\\
 BD+17$^\circ$2537	&	-	&	5900	&	$-$0.24 	&	4.00 	&	-	&	-	&	0.49 	&	0.56 	&	-	&	0.59 	&	0.22 	&	0.09 	&	0.39 	&	sgCH	&	8	\\
HD~127392	&	-	&	5600	&	$-$0.52 	&	3.90 	&	-	&	-	&	1.04 	&	0.48 	&	-	&	1.16 	&	0.70 	&	0.98 	&	0.87 	&	sgCH	&	8	\\
 BD$-$11$^\circ$3853	&	-	&	6000	&	$-$0.81 	&	4.30 	&	-	&	-	&	0.65 	&	0.48 	&	-	&	1.15 	&	1.10 	&	1.27 	&	0.93 	&	sgCH	&	8	\\
HD~141804  	&	-	&	6000	&	$-$0.41 	&	3.50 	&	-	&	-	&	0.60 	&	1.48 	&	1.25 	&	1.18 	&	1.10 	&	0.95 	&	1.09 	&	sgCH	&	8	\\
BD$-$10$^\circ$4311	&	-	&	5800	&	$-$0.58 	&	4.00 	&	-	&	-	&	0.85 	&	0.62 	&	-	&	0.70 	&	0.51 	&	0.94 	&	0.72 	&	sgCH	&	8	\\
CPD$-$62$^\circ$6195    	&	-	&	5400	&	$-$0.97 	&	3.50 	&	-	&	-	&	0.60 	&	1.57 	&	-	&	$-$0.09 	&	0.89 	&	1.16 	&	0.83 	&	sgCH	&	8	\\
HD~202020  	&	-	&	5600	&	$-$0.13 	&	4.00 	&	-	&	-	&	0.86 	&	0.31 	&	-	&	0.73 	&	0.09 	&	0.31 	&	0.46 	&	sgCH	&	8	\\
HD~224621  	&	-	&	6000	&	$-$0.41 	&	4.00 	&	-	&	-	&	0.91 	&	0.34 	&	-	&	0.63 	&	0.73 	&	1.09 	&	0.74 	&	sgCH	&	8	\\
HD~4395  	&	1.4	&	5550	&	$-$0.16 	&	3.70 	&	-	&	1.08 	&	0.65 	&	0.58 	&	0.79 	&	1.03 	&	0.42 	&	0.80 	&	0.76 	&	sgCH	&	9	\\
HD~48565 	&	1.2	&	6030	&	$-$0.59 	&	3.80 	&	-	&	1.73 	&	1.08 	&	0.90 	&	1.52 	&	1.46 	&	1.42 	&	1.51 	&	1.37 	&	sgCH	&	9	\\
HD~216219	&	1.6	&	5950	&	$-$0.17 	&	3.50 	&	-	&	1.80 	&	1.00 	&	0.98 	&	1.10 	&	1.04 	&	1.03 	&	0.99 	&	1.13 	&	sgCH	&	9	\\
HD~89668	&	-	&	5400	&	$-$0.13 	&	4.40 	&	0.05	&	1.06 	&	0.55 	&	-	&	$-$0.24 	&	1.87 	&	1.52 	&	1.44 	&	1.03 	&	sgCH	&	10	\\
HD~92545	&	1.2	&	6380	&	$-$0.21 	&	4.70 	&	0.68	&	-	&	0.23 	&	-	&	0.91 	&	0.95 	&	1.60 	&	-	&	0.92 	&	sgCH	&	10	\\
HD~122202	&	-	&	6430	&	$-$0.63 	&	4.00 	&	0.50	&	-	&	1.44 	&	-	&	0.33 	&	0.90 	&	1.62 	&	-	&	1.07 	&	sgCH	&	10	\\
HD~126681	&	-	&	5760	&	$-$0.90 	&	4.70 	&	-	&	-	&	0.02 	&	-	&	0.27 	&	-	&	0.67 	&	1.20 	&	0.54 	&	sgCH	&	10	\\
{\bf HD~164922}	&	-	&	5400	&	{\bf 0.22} 	&	4.30 	&	{\bf $-$0.15}	&	0.79 	&	0.14 	&	-	&	0.28 	&	0.15 	&	$-$0.09 	&	-	&	{\bf 0.25} 	&	sgCH	&	10	\\
HD~204613	&	1.1	&	5875	&	$-$0.24 	&	4.20 	&	0.49	&	1.71 	&	0.97 	&	1.14 	&	1.04 	&	1.21 	&	1.24 	&	1.02 	&	1.19 	&	sgCH	&	11	\\
HD~21922	&	-	&	5943	&	$-$0.48 	&	4.30 	&	0.27	&	0.25 	&	0.23 	&	0.33 	&	0.18 	&	-	&	0.34 	&	0.17 	&	0.25 	&	dwarf	&	11	\\
HD~36667	&	-	&	5776	&	$-$0.45 	&	4.00 	&	0.34	&	0.33 	&	0.38 	&	0.47 	&	0.33 	&	-	&	0.26 	&	0.40 	&	0.36 	&	dwarf	&	11	\\
HD~80218	&	-	&	6091	&	$-$0.28 	&	4.20 	&	0.21	&	0.33 	&	0.26 	&	0.00 	&	0.16 	&	-	&	$-$0.07 	&	0.04 	&	0.12 	&	dwarf	&	11	\\
HD~88446	&	-	&	5875	&	$-$0.35 	&	4.10 	&	0.34	&	0.57 	&	0.58 	&	0.66 	&	0.65 	&	-	&	0.58 	&	0.69 	&	0.62 	&	dwarf	&	11	\\
HD~140324	&	-	&	5822	&	$-$0.36 	&	4.10 	&	0.23	&	-	&	0.29 	&	0.18 	&	-	&	-	&	0.29 	&	-	&	0.25 	&	dwarf	&	11	\\
HD~220842	&	-	&	5761	&	$-$0.31 	&	4.20 	&	0.39	&	0.27 	&	0.35 	&	0.09 	&	0.27 	&	-	&	0.05 	&	0.12 	&	0.19 	&	dwarf	&	11	\\
HR~2906	&	-	&	6167	&	$-$0.18	&	4.09	&	-	&	-	&	0.17	&	0.16	&	0.29	&	-	&	-	&	0.16	&	0.20 	&	dwarf	&	12	\\
HR~4285	&	-	&	5890	&	$-$0.30	&	4.05	&	-	&	-	&	0.09	&	0.20	&	0.23	&	-	&	-	&	-	&	0.17 	&	dwarf	&	12	\\
HR~4395	&	-	&	6643	&	$-$0.10	&	3.98	&	-	&	-	&	-	&	-	&	0.57	&	-	&	-	&	-	&	0.57 	&	dwarf	&	12	\\
HR~5338	&	-	&	6177	&	$-$0.11	&	3.94	&	0.07	&	-	&	0.35	&	-	&	0.36	&	-	&	-	&	-	&	0.36 	&	dwarf	&	12	\\
HD~6434	&	-	&	5813	&	$-$0.54	&	4.42	&	-	&	-	&	0.28	&	0.34	&	0.34	&	-	&	-	&	0.16	&	0.28 	&	dwarf	&	12	\\
HD~15306 & -    &   6750     &  $-$0.50  &   4.05   &0.47   & 0.76  &   0.89    &   0.77     & 0.71       & -   &   0.72   &  0.83   &   0.78      & dwarf   & 13 \\
HD~202400 & -   &   6200   &    $-$0.70   &  4.00    & -     & 0.86  &   1.15    & 1.29        &  1.01   &  -    &  1.34    &1.48      &     1.19  &  dwarf    &  13  \\
HD~221531	&	-	&	6460	&$	-0.30	$&	4.24	&	0.25	&	0.60	&	0.78	&	0.67	&	0.34	&	-	&	0.31	&	0.64	&	0.56 	&	dwarf	&	13	\\
HD~147609	&	-	&	6270	&$	-0.50	$&	3.50	&	0.29	&	-	&	0.96	&	0.80	&	-	&	-	&	-	&	0.98	&	0.91 	&    sgCH		&	13	\\
HD~177645	&	-	&	7150	&$	-0.20	$&	3.55	&	-	&	-	&	1.52	&	1.36	&	-	&	-	&	-	&	0.96	&	1.28 	&	sgCH	&	13	\\

\hline
\multicolumn{16}{l}{$1$ \citealt{2006Allen}; $2$ \citealt{2011Pereira}; $3$ \citealt{1997Porto}; $4$ \citealt{2007Gray}; $5$ \citealt{2005Pereira}; $6$ \citealt{2003Pereira};  }\\

\multicolumn{16}{l}{$7$ \citealt{1993ApJ}; $8$ \citealt{1991LuckL}; $ 9 $ \citealt{2014Karinkuzh}; $10$ \citealt{2015Karinkuzhi}; $11$\citealt{2003MNRAS}; $12$ \citealt{1993Edvardsson} }\\
\multicolumn{16}{l}{$13$ \citealt{1994North}} \\
\end{tabular}

 }

\end{table*}

\section*{Acknowledgements}
We are thankful to the referee for his constructive comments and suggestions that lead to improve the manuscript. We thank Kefeng Tan for giving access to the model atmosphere softwares and Yujuan Liu for stimulating discussions. This study was supported by the National Natural Science Foundation of China under grant number 11390371, 11233004, U1431106 and 11550110492. Y.B.K thanks the Chinese \mbox{Academy} of Sciences for support through CAS PIFI fellowship.







\begin{thebibliography}{}
\makeatletter
\relax
\def\mn@urlcharsother{\let\do\@makeother \do\$\do\&\do\#\do\^\do\_\do\%\do\~}
\def\mn@doi{\begingroup\mn@urlcharsother \@ifnextchar [ {\mn@doi@}
  {\mn@doi@[]}}
\def\mn@doi@[#1]#2{\def\@tempa{#1}\ifx\@tempa\@empty \href
  {http://dx.doi.org/#2} {doi:#2}\else \href {http://dx.doi.org/#2} {#1}\fi
  \endgroup}
\def\mn@eprint#1#2{\mn@eprint@#1:#2::\@nil}
\def\mn@eprint@arXiv#1{\href {http://arxiv.org/abs/#1} {{\tt arXiv:#1}}}
\def\mn@eprint@dblp#1{\href {http://dblp.uni-trier.de/rec/bibtex/#1.xml}
  {dblp:#1}}
\def\mn@eprint@#1:#2:#3:#4\@nil{\def\@tempa {#1}\def\@tempb {#2}\def\@tempc
  {#3}\ifx \@tempc \@empty \let \@tempc \@tempb \let \@tempb \@tempa \fi \ifx
  \@tempb \@empty \def\@tempb {arXiv}\fi \@ifundefined
  {mn@eprint@\@tempb}{\@tempb:\@tempc}{\expandafter \expandafter \csname
  mn@eprint@\@tempb\endcsname \expandafter{\@tempc}}}

\bibitem[\protect\citeauthoryear{{Allen} \& {Barbuy}}{{Allen} \&
  {Barbuy}}{2006}]{2006Allen}
{Allen} D.~M.,  {Barbuy} B.,  2006, \mn@doi [\aap]
  {10.1051/0004-6361:20064912}, \href
  {http://ads.bao.ac.cn/abs/2006A%26A...454..895A} {454, 895}

\bibitem[\protect\citeauthoryear{{Allen} \& {Porto de Mello}}{{Allen} \& {Porto
  de Mello}}{2011}]{Allen2011}
{Allen} D.~M.,  {Porto de Mello} G.~F.,  2011, \mn@doi [\aap]
  {10.1051/0004-6361/200912356}, \href
  {http://ads.bao.ac.cn/abs/2011A%26A...525A..63A} {525, A63}

\bibitem[\protect\citeauthoryear{{Allende Prieto} \& {Lambert}}{{Allende
  Prieto} \& {Lambert}}{1999}]{1999Prieto}
{Allende Prieto} C.,  {Lambert} D.~L.,  1999, \aap, \href
  {http://cdsads.u-strasbg.fr/abs/1999A%26A...352..555A} {352, 555}

\bibitem[\protect\citeauthoryear{{Alonso}, {Arribas}  \&
  {Martinez-Roger}}{{Alonso} et~al.}{1995}]{Alonso1995}
{Alonso} A.,  {Arribas} S.,   {Martinez-Roger} C.,  1995, \aap, \href
  {http://ads.bao.ac.cn/abs/1995A%26A...297..197A} {297, 197}

\bibitem[\protect\citeauthoryear{{Alonso}, {Arribas}  \&
  {Martinez-Roger}}{{Alonso} et~al.}{1996}]{Alonso1996}
{Alonso} A.,  {Arribas} S.,   {Martinez-Roger} C.,  1996, \aap, \href
  {http://ads.bao.ac.cn/abs/1996A%26A...313..873A} {313, 873}

\bibitem[\protect\citeauthoryear{{Bard} \& {Kock}}{{Bard} \&
  {Kock}}{1994}]{Bard1994}
{Bard} A.,  {Kock} M.,  1994, \aap, \href
  {http://ads.bao.ac.cn/abs/1994A%26A...282.1014B} {282, 1014}

\bibitem[\protect\citeauthoryear{{Bard}, {Kock}  \& {Kock}}{{Bard}
  et~al.}{1991}]{BardKock1991}
{Bard} A.,  {Kock} A.,   {Kock} M.,  1991, \aap, \href
  {http://ads.bao.ac.cn/abs/1991A%26A...248..315B} {248, 315}

\bibitem[\protect\citeauthoryear{{Barstow}, {Boyce}, {Welsh}, {Lallement},
  {Barstow}, {Forbes}  \& {Preval}}{{Barstow} et~al.}{2010}]{2010Barstow}
{Barstow} M.~A.,  {Boyce} D.~D.,  {Welsh} B.~Y.,  {Lallement} R.,  {Barstow}
  J.~K.,  {Forbes} A.~E.,   {Preval} S.,  2010, \mn@doi [\apj]
  {10.1088/0004-637X/723/2/1762}, \href
  {http://ads.bao.ac.cn/abs/2010ApJ...723.1762B} {723, 1762}

\bibitem[\protect\citeauthoryear{{Beers}, {Drilling}, {Rossi}, {Chiba}, {Rhee},
  {F{\"u}hrmeister}, {Norris}  \& {von Hippel}}{{Beers}
  et~al.}{2002}]{2002Beers}
{Beers} T.~C.,  {Drilling} J.~S.,  {Rossi} S.,  {Chiba} M.,  {Rhee} J.,
  {F{\"u}hrmeister} B.,  {Norris} J.~E.,   {von Hippel} T.,  2002, \mn@doi
  [\aj] {10.1086/341377}, \href {http://ads.bao.ac.cn/abs/2002AJ....124..931B}
  {124, 931}

\bibitem[\protect\citeauthoryear{{Bensby}, {Feltzing}  \& {Oey}}{{Bensby}
  et~al.}{2014}]{Bensby2014}
{Bensby} T.,  {Feltzing} S.,   {Oey} M.~S.,  2014, \mn@doi [\aap]
  {10.1051/0004-6361/201322631}, \href
  {http://ads.bao.ac.cn/abs/2014A%26A...562A..71B} {562, A71}

\bibitem[\protect\citeauthoryear{{Bergeat} \& {Knapik}}{{Bergeat} \&
  {Knapik}}{1997}]{1997Bergeat}
{Bergeat} J.,  {Knapik} A.,  1997, \aap, \href
  {http://ads.bao.ac.cn/abs/1997A%26A...321L...9B} {321, L9}

\bibitem[\protect\citeauthoryear{{Bidelman}}{{Bidelman}}{1981}]{1981Bidelman}
{Bidelman} W.~P.,  1981, \mn@doi [\aj] {10.1086/112913}, \href
  {http://ads.bao.ac.cn/abs/1981AJ.....86..553B} {86, 553}

\bibitem[\protect\citeauthoryear{{Bidelman}}{{Bidelman}}{1983}]{1983Bidelman}
{Bidelman} W.~P.,  1983, \mn@doi [\aj] {10.1086/113408}, \href
  {http://ads.bao.ac.cn/abs/1983AJ.....88.1182B} {88, 1182}

\bibitem[\protect\citeauthoryear{{Bidelman}}{{Bidelman}}{1985}]{1985Bidelman}
{Bidelman} W.~P.,  1985, \mn@doi [\aj] {10.1086/113737}, \href
  {http://ads.bao.ac.cn/abs/1985AJ.....90..341B} {90, 341}

\bibitem[\protect\citeauthoryear{{Bidelman} \& {Keenan}}{{Bidelman} \&
  {Keenan}}{1951}]{Bidelman1951}
{Bidelman} W.~P.,  {Keenan} P.~C.,  1951, \mn@doi [\apj] {10.1086/145488},
  \href {http://ads.bao.ac.cn/abs/1951ApJ...114..473B} {114, 473}

\bibitem[\protect\citeauthoryear{{Blackwell}, {Petford}, {Shallis}  \&
  {Simmons}}{{Blackwell} et~al.}{1982a}]{Blackwell1982b}
{Blackwell} D.~E.,  {Petford} A.~D.,  {Shallis} M.~J.,   {Simmons} G.~J.,
  1982a, \mn@doi [\mnras] {10.1093/mnras/199.1.43}, \href
  {http://ads.bao.ac.cn/abs/1982MNRAS.199...43B} {199, 43}

\bibitem[\protect\citeauthoryear{{Blackwell}, {Petford}  \&
  {Simmons}}{{Blackwell} et~al.}{1982b}]{Blackwell1982c}
{Blackwell} D.~E.,  {Petford} A.~D.,   {Simmons} G.~J.,  1982b, \mn@doi
  [\mnras] {10.1093/mnras/201.3.595}, \href
  {http://ads.bao.ac.cn/abs/1982MNRAS.201..595B} {201, 595}

\bibitem[\protect\citeauthoryear{{Boffin} \& {Jorissen}}{{Boffin} \&
  {Jorissen}}{1988}]{Boffin1988}
{Boffin} H.~M.~J.,  {Jorissen} A.,  1988, \aap, \href
  {http://ads.bao.ac.cn/abs/1988A%26A...205..155B} {205, 155}

\bibitem[\protect\citeauthoryear{{B{\"o}hm-Vitense}, {Carpenter}, {Robinson},
  {Ake}  \& {Brown}}{{B{\"o}hm-Vitense} et~al.}{2000}]{Vitense2000}
{B{\"o}hm-Vitense} E.,  {Carpenter} K.,  {Robinson} R.,  {Ake} T.,   {Brown}
  J.,  2000, \mn@doi [\apj] {10.1086/308678}, \href
  {http://ads.bao.ac.cn/abs/2000ApJ...533..969B} {533, 969}

\bibitem[\protect\citeauthoryear{{Boothroyd} \& {Sackmann}}{{Boothroyd} \&
  {Sackmann}}{1992}]{1992Boothroyd}
{Boothroyd} A.~I.,  {Sackmann} I.-J.,  1992, \mn@doi [\apjl] {10.1086/186441},
  \href {http://ads.bao.ac.cn/abs/1992ApJ...393L..21B} {393, L21}

\bibitem[\protect\citeauthoryear{{Catal{\'a}n}, {Isern}, {Garc{\'{\i}}a-Berro}
  \& {Ribas}}{{Catal{\'a}n} et~al.}{2008}]{2008Catal}
{Catal{\'a}n} S.,  {Isern} J.,  {Garc{\'{\i}}a-Berro} E.,   {Ribas} I.,  2008,
  \mn@doi [\mnras] {10.1111/j.1365-2966.2008.13356.x}, \href
  {http://ads.bao.ac.cn/abs/2008MNRAS.387.1693C} {387, 1693}

\bibitem[\protect\citeauthoryear{{Colucci}, {Bernstein}, {Cameron}  \&
  {McWilliam}}{{Colucci} et~al.}{2012}]{Colucci2012}
{Colucci} J.~E.,  {Bernstein} R.~A.,  {Cameron} S.~A.,   {McWilliam} A.,  2012,
  \mn@doi [\apj] {10.1088/0004-637X/746/1/29}, \href
  {http://ads.bao.ac.cn/abs/2012ApJ...746...29C} {746, 29}

\bibitem[\protect\citeauthoryear{{Covey} et~al.,}{{Covey}
  et~al.}{2016}]{2016Covey}
{Covey} K.~R.,  et~al., 2016, \mn@doi [\apj] {10.3847/0004-637X/822/2/81},
  \href {http://ads.bao.ac.cn/abs/2016ApJ...822...81C} {822, 81}

\bibitem[\protect\citeauthoryear{{Cutri} et~al.,}{{Cutri}
  et~al.}{2003}]{2003Cutri}
{Cutri} R.~M.,  et~al., 2003, VizieR Online Data Catalog, \href
  {http://adsabs.harvard.edu/abs/2003yCat.2246....0C} {2246}

\bibitem[\protect\citeauthoryear{{da Silva}, {Milone}  \& {Rocha-Pinto}}{{da
  Silva} et~al.}{2015}]{Silva2015}
{da Silva} R.,  {Milone} A.~d.~C.,   {Rocha-Pinto} H.~J.,  2015, \mn@doi [\aap]
  {10.1051/0004-6361/201525770}, \href
  {http://ads.bao.ac.cn/abs/2015A%26A...580A..24D} {580, A24}

\bibitem[\protect\citeauthoryear{{de Castro}, {Pereira}, {Roig}, {Jilinski},
  {Drake}, {Chavero}  \& {Sales Silva}}{{de Castro} et~al.}{2016}]{Castro2016}
{de Castro} D.~B.,  {Pereira} C.~B.,  {Roig} F.,  {Jilinski} E.,  {Drake}
  N.~A.,  {Chavero} C.,   {Sales Silva} J.~V.,  2016, \mn@doi [\mnras]
  {10.1093/mnras/stw815}, \href {http://ads.bao.ac.cn/abs/2016MNRAS.459.4299D}
  {459, 4299}

\bibitem[\protect\citeauthoryear{{Dominguez}, {Chieffi}, {Limongi}  \&
  {Straniero}}{{Dominguez} et~al.}{1999}]{1999Dominguez}
{Dominguez} I.,  {Chieffi} A.,  {Limongi} M.,   {Straniero} O.,  1999, \mn@doi
  [\apj] {10.1086/307787}, \href {http://ads.bao.ac.cn/abs/1999ApJ...524..226D}
  {524, 226}

\bibitem[\protect\citeauthoryear{{Edvardsson}, {Andersen}, {Gustafsson},
  {Lambert}, {Nissen}  \& {Tomkin}}{{Edvardsson} et~al.}{1993}]{1993Edvardsson}
{Edvardsson} B.,  {Andersen} J.,  {Gustafsson} B.,  {Lambert} D.~L.,  {Nissen}
  P.~E.,   {Tomkin} J.,  1993, \aap, \href
  {http://ads.bao.ac.cn/abs/1993A%26A...275..101E} {275, 101}

\bibitem[\protect\citeauthoryear{{Escorza} et~al.,}{{Escorza}
  et~al.}{2017}]{2017Escorza}
{Escorza} A.,  et~al., 2017, preprint, \href
  {http://adsabs.harvard.edu/abs/2017arXiv171002029E} {} (\mn@eprint {arXiv}
  {1710.02029})

\bibitem[\protect\citeauthoryear{{Gaia Collaboration} et~al.,}{{Gaia
  Collaboration} et~al.}{2016}]{2016Gaia}
{Gaia Collaboration} et~al., 2016, \mn@doi [\aap]
  {10.1051/0004-6361/201629512}, \href
  {http://ads.bao.ac.cn/abs/2016A%26A...595A...2G} {595, A2}

\bibitem[\protect\citeauthoryear{{Gianninas}, {Bergeron}  \&
  {Ruiz}}{{Gianninas} et~al.}{2011}]{2011Gianninas}
{Gianninas} A.,  {Bergeron} P.,   {Ruiz} M.~T.,  2011, \mn@doi [\apj]
  {10.1088/0004-637X/743/2/138}, \href
  {http://ads.bao.ac.cn/abs/2011ApJ...743..138G} {743, 138}

\bibitem[\protect\citeauthoryear{{Gray} \& {Griffin}}{{Gray} \&
  {Griffin}}{2007}]{2007Gray}
{Gray} R.~O.,  {Griffin} R.~E.~M.,  2007, \mn@doi [\aj] {10.1086/518476}, \href
  {http://ads.bao.ac.cn/abs/2007AJ....134...96G} {134, 96}

\bibitem[\protect\citeauthoryear{{Gray}, {McGahee}, {Griffin}  \&
  {Corbally}}{{Gray} et~al.}{2011}]{Gray2011}
{Gray} R.~O.,  {McGahee} C.~E.,  {Griffin} R.~E.~M.,   {Corbally} C.~J.,  2011,
  \mn@doi [\aj] {10.1088/0004-6256/141/5/160}, \href
  {http://ads.bao.ac.cn/abs/2011AJ....141..160G} {141, 160}

\bibitem[\protect\citeauthoryear{{Grevesse} \& {Sauval}}{{Grevesse} \&
  {Sauval}}{1998}]{Grevesse1998}
{Grevesse} N.,  {Sauval} A.~J.,  1998, in {Fr{\"o}hlich} C.,  {Huber} M.~C.~E.,
   {Solanki} S.~K.,   {von Steiger} R.,  eds, Solar Composition and Its
  Evolution -- From Core to Corona. p.~161

\bibitem[\protect\citeauthoryear{{Guillout} et~al.,}{{Guillout}
  et~al.}{2009}]{2009Guillout}
{Guillout} P.,  et~al., 2009, \mn@doi [\aap] {10.1051/0004-6361/200811313},
  \href {http://ads.bao.ac.cn/abs/2009A%26A...504..829G} {504, 829}

\bibitem[\protect\citeauthoryear{{Han}, {Eggleton}, {Podsiadlowski}  \&
  {Tout}}{{Han} et~al.}{1995}]{Han1995}
{Han} Z.,  {Eggleton} P.~P.,  {Podsiadlowski} P.,   {Tout} C.~A.,  1995,
  \mn@doi [\mnras] {10.1093/mnras/277.4.1443}, \href
  {http://ads.bao.ac.cn/abs/1995MNRAS.277.1443H} {277, 1443}

\bibitem[\protect\citeauthoryear{{Holberg}, {Oswalt}, {Sion}, {Barstow}  \&
  {Burleigh}}{{Holberg} et~al.}{2013}]{Holberg2013}
{Holberg} J.~B.,  {Oswalt} T.~D.,  {Sion} E.~M.,  {Barstow} M.~A.,   {Burleigh}
  M.~R.,  2013, \mn@doi [\mnras] {10.1093/mnras/stt1433}, \href
  {http://adsabs.harvard.edu/abs/2013MNRAS.435.2077H} {435, 2077}

\bibitem[\protect\citeauthoryear{{Hurley}, {Pols}  \& {Tout}}{{Hurley}
  et~al.}{2000}]{Hurley2000}
{Hurley} J.~R.,  {Pols} O.~R.,   {Tout} C.~A.,  2000, \mn@doi [\mnras]
  {10.1046/j.1365-8711.2000.03426.x}, \href
  {http://ads.bao.ac.cn/abs/2000MNRAS.315..543H} {315, 543}

\bibitem[\protect\citeauthoryear{{Johnson}, {McWilliam}  \& {Rich}}{{Johnson}
  et~al.}{2013}]{2013Johnson}
{Johnson} C.~I.,  {McWilliam} A.,   {Rich} R.~M.,  2013, \mn@doi [\apjl]
  {10.1088/2041-8205/775/1/L27}, \href
  {http://ads.bao.ac.cn/abs/2013ApJ...775L..27J} {775, L27}

\bibitem[\protect\citeauthoryear{{Jorissen}, {Van Eck}, {Mayor}  \&
  {Udry}}{{Jorissen} et~al.}{1998}]{Jorissen1998}
{Jorissen} A.,  {Van Eck} S.,  {Mayor} M.,   {Udry} S.,  1998, \aap, \href
  {http://ads.bao.ac.cn/abs/1998A%26A...332..877J} {332, 877}

\bibitem[\protect\citeauthoryear{{Karakas} \& {Lugaro}}{{Karakas} \&
  {Lugaro}}{2016}]{2016Karakas}
{Karakas} A.~I.,  {Lugaro} M.,  2016, \mn@doi [\apj]
  {10.3847/0004-637X/825/1/26}, \href
  {http://ads.bao.ac.cn/abs/2016ApJ...825...26K} {825, 26}

\bibitem[\protect\citeauthoryear{{Karinkuzhi} \& {Goswami}}{{Karinkuzhi} \&
  {Goswami}}{2014}]{2014Karinkuzh}
{Karinkuzhi} D.,  {Goswami} A.,  2014, \mn@doi [\mnras] {10.1093/mnras/stu148},
  \href {http://ads.bao.ac.cn/abs/2014MNRAS.440.1095K} {440, 1095}

\bibitem[\protect\citeauthoryear{{Karinkuzhi} \& {Goswami}}{{Karinkuzhi} \&
  {Goswami}}{2015}]{2015Karinkuzhi}
{Karinkuzhi} D.,  {Goswami} A.,  2015, \mn@doi [\mnras]
  {10.1093/mnras/stu2079}, \href {http://ads.bao.ac.cn/abs/2015MNRAS.446.2348K}
  {446, 2348}

\bibitem[\protect\citeauthoryear{{Kawka} \& {Vennes}}{{Kawka} \&
  {Vennes}}{2010}]{2010Kawka}
{Kawka} A.,  {Vennes} S.,  2010, in {Pr{\v s}a} A.,  {Zejda} M.,  eds,
  Astronomical Society of the Pacific Conference Series Vol. 435, Binaries -
  Key to Comprehension of the Universe. p.~189

\bibitem[\protect\citeauthoryear{{Kiraga}}{{Kiraga}}{2012}]{2012Kiraga}
{Kiraga} M.,  2012, \actaa, \href
  {http://adsabs.harvard.edu/abs/2012AcA....62...67K} {62, 67}

\bibitem[\protect\citeauthoryear{{Kurucz}}{{Kurucz}}{1993}]{1993Kur}
{Kurucz} R.,  1993, ATLAS9 Stellar Atmosphere Programs and 2 km/s grid.~Kurucz
  CD-ROM No.~13.~ Cambridge, Mass.: Smithsonian Astrophysical Observatory,
  1993., \href {http://ads.bao.ac.cn/abs/1993KurCD..13.....K} {13}

\bibitem[\protect\citeauthoryear{{Lambert}, {Smith}  \& {Heath}}{{Lambert}
  et~al.}{1993}]{1993Lambert}
{Lambert} D.~L.,  {Smith} V.~V.,   {Heath} J.,  1993, \mn@doi [\pasp]
  {10.1086/133195}, \href {http://ads.bao.ac.cn/abs/1993PASP..105..568L} {105,
  568}

\bibitem[\protect\citeauthoryear{{Liu}, {Zhao}, {Shi}, {Pietrzy{\'n}ski}  \&
  {Gieren}}{{Liu} et~al.}{2007}]{2007Liu}
{Liu} Y.~J.,  {Zhao} G.,  {Shi} J.~R.,  {Pietrzy{\'n}ski} G.,   {Gieren} W.,
  2007, \mn@doi [\mnras] {10.1111/j.1365-2966.2007.11852.x}, \href
  {http://ads.bao.ac.cn/abs/2007MNRAS.382..553L} {382, 553}

\bibitem[\protect\citeauthoryear{{Liu}, {Stancliffe}, {Abate}  \&
  {Matrozis}}{{Liu} et~al.}{2017}]{2017arXivliu}
{Liu} Z.-W.,  {Stancliffe} R.~J.,  {Abate} C.,   {Matrozis} E.,  2017, \mn@doi
  [\apj] {10.3847/1538-4357/aa8622}, \href
  {http://adsabs.harvard.edu/abs/2017ApJ...846..117L} {846, 117}

\bibitem[\protect\citeauthoryear{{Lu}, {Dawson}, {Upgren}  \& {Weis}}{{Lu}
  et~al.}{1983}]{1983Lu}
{Lu} P.~K.,  {Dawson} D.~W.,  {Upgren} A.~R.,   {Weis} E.~W.,  1983, \mn@doi
  [\apjs] {10.1086/190863}, \href
  {http://ads.bao.ac.cn/abs/1983ApJS...52..169L} {52, 169}

\bibitem[\protect\citeauthoryear{{Luck} \& {Bond}}{{Luck} \&
  {Bond}}{1991}]{1991LuckL}
{Luck} R.~E.,  {Bond} H.~E.,  1991, \mn@doi [\apjs] {10.1086/191615}, \href
  {http://ads.bao.ac.cn/abs/1991ApJS...77..515L} {77, 515}

\bibitem[\protect\citeauthoryear{{Mamajek}, {Meyer}  \& {Liebert}}{{Mamajek}
  et~al.}{2002}]{Mamajek2002}
{Mamajek} E.~E.,  {Meyer} M.~R.,   {Liebert} J.,  2002, \mn@doi [\aj]
  {10.1086/341952}, \href {http://ads.bao.ac.cn/abs/2002AJ....124.1670M} {124,
  1670}

\bibitem[\protect\citeauthoryear{{McCall}}{{McCall}}{2004}]{McCall2004}
{McCall} M.~L.,  2004, \mn@doi [\aj] {10.1086/424933}, \href
  {http://ads.bao.ac.cn/abs/2004AJ....128.2144M} {128, 2144}

\bibitem[\protect\citeauthoryear{{McClure}}{{McClure}}{1984}]{McClure1984}
{McClure} R.~D.,  1984, \mn@doi [\pasp] {10.1086/131310}, \href
  {http://ads.bao.ac.cn/abs/1984PASP...96..117M} {96, 117}

\bibitem[\protect\citeauthoryear{{McClure} \& {Woodsworth}}{{McClure} \&
  {Woodsworth}}{1990}]{1990McClure}
{McClure} R.~D.,  {Woodsworth} A.~W.,  1990, \mn@doi [\apj] {10.1086/168573},
  \href {http://ads.bao.ac.cn/abs/1990ApJ...352..709M} {352, 709}

\bibitem[\protect\citeauthoryear{{McClure}, {Fletcher}  \& {Nemec}}{{McClure}
  et~al.}{1980}]{McClure1980}
{McClure} R.~D.,  {Fletcher} J.~M.,   {Nemec} J.~M.,  1980, \mn@doi [\apjl]
  {10.1086/183252}, \href {http://ads.bao.ac.cn/abs/1980ApJ...238L..35M} {238,
  L35}

\bibitem[\protect\citeauthoryear{{Mel{\'e}ndez} et~al.,}{{Mel{\'e}ndez}
  et~al.}{2014}]{Melendez2014}
{Mel{\'e}ndez} J.,  et~al., 2014, \mn@doi [\apj] {10.1088/0004-637X/791/1/14},
  \href {http://ads.bao.ac.cn/abs/2014ApJ...791...14M} {791, 14}

\bibitem[\protect\citeauthoryear{{Merle}, {Jorissen}, {Van Eck}, {Masseron}  \&
  {Van Winckel}}{{Merle} et~al.}{2016}]{Merle2016}
{Merle} T.,  {Jorissen} A.,  {Van Eck} S.,  {Masseron} T.,   {Van Winckel} H.,
  2016, \mn@doi [\aap] {10.1051/0004-6361/201526944}, \href
  {http://ads.bao.ac.cn/abs/2016A%26A...586A.151M} {586, A151}

\bibitem[\protect\citeauthoryear{{Montes} \& {Ramsey}}{{Montes} \&
  {Ramsey}}{1998}]{1998Montes}
{Montes} D.,  {Ramsey} L.~W.,  1998, \aap, \href
  {http://adsabs.harvard.edu/abs/1998A%26A...340L...5M} {340, L5}

\bibitem[\protect\citeauthoryear{{North}, {Berthet}  \& {Lanz}}{{North}
  et~al.}{1994}]{1994North}
{North} P.,  {Berthet} S.,   {Lanz} T.,  1994, \aap, \href
  {http://ads.bao.ac.cn/abs/1994A%26A...281..775N} {281, 775}

\bibitem[\protect\citeauthoryear{{North}, {Jorissen}  \& {Mayor}}{{North}
  et~al.}{2000}]{2000North}
{North} P.,  {Jorissen} A.,   {Mayor} M.,  2000, in {Wing} R.~F.,  ed.,  IAU
  Symposium Vol. 177, The Carbon Star Phenomenon. p.~269

\bibitem[\protect\citeauthoryear{{O'Brian}, {Wickliffe}, {Lawler}, {Whaling}
  \& {Brault}}{{O'Brian} et~al.}{1991}]{OBrian1991}
{O'Brian} T.~R.,  {Wickliffe} M.~E.,  {Lawler} J.~E.,  {Whaling} W.,   {Brault}
  J.~W.,  1991, \mn@doi [Journal of the Optical Society of America B Optical
  Physics] {10.1364/JOSAB.8.001185}, \href
  {http://ads.bao.ac.cn/abs/1991JOSAB...8.1185O} {8, 1185}

\bibitem[\protect\citeauthoryear{{Paxton}, {Bildsten}, {Dotter}, {Herwig},
  {Lesaffre}  \& {Timmes}}{{Paxton} et~al.}{2011}]{2011Paxton}
{Paxton} B.,  {Bildsten} L.,  {Dotter} A.,  {Herwig} F.,  {Lesaffre} P.,
  {Timmes} F.,  2011, \mn@doi [\apjs] {10.1088/0067-0049/192/1/3}, \href
  {http://ads.bao.ac.cn/abs/2011ApJS..192....3P} {192, 3}

\bibitem[\protect\citeauthoryear{{Paxton} et~al.,}{{Paxton}
  et~al.}{2013}]{2013Paxton}
{Paxton} B.,  et~al., 2013, \mn@doi [\apjs] {10.1088/0067-0049/208/1/4}, \href
  {http://ads.bao.ac.cn/abs/2013ApJS..208....4P} {208, 4}

\bibitem[\protect\citeauthoryear{{Paxton} et~al.,}{{Paxton}
  et~al.}{2015}]{2015Paxton}
{Paxton} B.,  et~al., 2015, \mn@doi [\apjs] {10.1088/0067-0049/220/1/15}, \href
  {http://ads.bao.ac.cn/abs/2015ApJS..220...15P} {220, 15}

\bibitem[\protect\citeauthoryear{{Pereira}}{{Pereira}}{2005}]{2005Pereira}
{Pereira} C.~B.,  2005, \mn@doi [\aj] {10.1086/428755}, \href
  {http://ads.bao.ac.cn/abs/2005AJ....129.2469P} {129, 2469}

\bibitem[\protect\citeauthoryear{{Pereira} \& {Drake}}{{Pereira} \&
  {Drake}}{2011}]{2011Pereira}
{Pereira} C.~B.,  {Drake} N.~A.,  2011, \mn@doi [\aj]
  {10.1088/0004-6256/141/3/79}, \href
  {http://ads.bao.ac.cn/abs/2011AJ....141...79P} {141, 79}

\bibitem[\protect\citeauthoryear{{Pereira} \& {Junqueira}}{{Pereira} \&
  {Junqueira}}{2003}]{2003Pereira}
{Pereira} C.~B.,  {Junqueira} S.,  2003, \mn@doi [\aap]
  {10.1051/0004-6361:20030209}, \href
  {http://ads.bao.ac.cn/abs/2003A%26A...402.1061P} {402, 1061}

\bibitem[\protect\citeauthoryear{{Pereira}, {Sales Silva}, {Chavero}, {Roig}
  \& {Jilinski}}{{Pereira} et~al.}{2011}]{Pereira2011}
{Pereira} C.~B.,  {Sales Silva} J.~V.,  {Chavero} C.,  {Roig} F.,   {Jilinski}
  E.,  2011, \mn@doi [\aap] {10.1051/0004-6361/201117070}, \href
  {http://ads.bao.ac.cn/abs/2011A%26A...533A..51P} {533, A51}

\bibitem[\protect\citeauthoryear{{Pomp{\'e}ia} \& {Allen}}{{Pomp{\'e}ia} \&
  {Allen}}{2008}]{2008Pomp}
{Pomp{\'e}ia} L.,  {Allen} D.~M.,  2008, \mn@doi [\aap]
  {10.1051/0004-6361:200809707}, \href
  {http://ads.bao.ac.cn/abs/2008A%26A...488..723P} {488, 723}

\bibitem[\protect\citeauthoryear{{Porto de Mello} \& {da Silva}}{{Porto de
  Mello} \& {da Silva}}{1997a}]{1997Porto}
{Porto de Mello} G.~F.,  {da Silva} L.,  1997a, \mn@doi [\apjl]
  {10.1086/310504}, \href {http://ads.bao.ac.cn/abs/1997ApJ...476L..89P} {476,
  L89}

\bibitem[\protect\citeauthoryear{{Porto de Mello} \& {da Silva}}{{Porto de
  Mello} \& {da Silva}}{1997b}]{Porto1997}
{Porto de Mello} G.~F.,  {da Silva} L.,  1997b, \mn@doi [\apjl]
  {10.1086/310504}, \href {http://ads.bao.ac.cn/abs/1997ApJ...476L..89P} {476,
  L89}

\bibitem[\protect\citeauthoryear{{Ram{\'{\i}}rez} \&
  {Mel{\'e}ndez}}{{Ram{\'{\i}}rez} \& {Mel{\'e}ndez}}{2004}]{Ramirez2004}
{Ram{\'{\i}}rez} I.,  {Mel{\'e}ndez} J.,  2004, \mn@doi [\apj]
  {10.1086/421041}, \href {http://ads.bao.ac.cn/abs/2004ApJ...609..417R} {609,
  417}

\bibitem[\protect\citeauthoryear{{Reddy}, {Tomkin}, {Lambert}  \& {Allende
  Prieto}}{{Reddy} et~al.}{2003}]{2003MNRAS}
{Reddy} B.~E.,  {Tomkin} J.,  {Lambert} D.~L.,   {Allende Prieto} C.,  2003,
  \mn@doi [\mnras] {10.1046/j.1365-8711.2003.06305.x}, \href
  {http://ads.bao.ac.cn/abs/2003MNRAS.340..304R} {340, 304}

\bibitem[\protect\citeauthoryear{{Reiners}, {Sch{\"u}ssler}  \&
  {Passegger}}{{Reiners} et~al.}{2014}]{2014Reiners}
{Reiners} A.,  {Sch{\"u}ssler} M.,   {Passegger} V.~M.,  2014, \mn@doi [\apj]
  {10.1088/0004-637X/794/2/144}, \href
  {http://adsabs.harvard.edu/abs/2014ApJ...794..144R} {794, 144}

\bibitem[\protect\citeauthoryear{{Reyniers}, {Van Winckel}, {Bi{\'e}mont}  \&
  {Quinet}}{{Reyniers} et~al.}{2002}]{2002Reyniers}
{Reyniers} M.,  {Van Winckel} H.,  {Bi{\'e}mont} E.,   {Quinet} P.,  2002,
  \mn@doi [\aap] {10.1051/0004-6361:20021502}, \href
  {http://ads.bao.ac.cn/abs/2002A%26A...395L..35R} {395, L35}

\bibitem[\protect\citeauthoryear{{Rojas}, {Drake}, {Pereira}  \&
  {Kholtygin}}{{Rojas} et~al.}{2013}]{Rojas2013}
{Rojas} M.,  {Drake} N.~A.,  {Pereira} C.~B.,   {Kholtygin} A.~F.,  2013,
  \mn@doi [Astrophysics] {10.1007/s10511-013-9267-8}, \href
  {http://ads.bao.ac.cn/abs/2013Ap.....56...57R} {56, 57}

\bibitem[\protect\citeauthoryear{{Schlegel}, {Finkbeiner}  \&
  {Davis}}{{Schlegel} et~al.}{1998}]{Schlegel1998}
{Schlegel} D.~J.,  {Finkbeiner} D.~P.,   {Davis} M.,  1998, \mn@doi [\apj]
  {10.1086/305772}, \href {http://ads.bao.ac.cn/abs/1998ApJ...500..525S} {500,
  525}

\bibitem[\protect\citeauthoryear{{Smith} \& {Lambert}}{{Smith} \&
  {Lambert}}{1986}]{1986Smith}
{Smith} V.~V.,  {Lambert} D.~L.,  1986, \mn@doi [\apj] {10.1086/164068}, \href
  {http://ads.bao.ac.cn/abs/1986ApJ...303..226S} {303, 226}

\bibitem[\protect\citeauthoryear{{Smith}, {Coleman}  \& {Lambert}}{{Smith}
  et~al.}{1993}]{1993ApJ}
{Smith} V.~V.,  {Coleman} H.,   {Lambert} D.~L.,  1993, \mn@doi [\apj]
  {10.1086/173311}, \href {http://ads.bao.ac.cn/abs/1993ApJ...417..287S} {417,
  287}

\bibitem[\protect\citeauthoryear{{Sneden}, {Pilachowski}  \&
  {Lambert}}{{Sneden} et~al.}{1981}]{Sneden1981}
{Sneden} C.,  {Pilachowski} C.~A.,   {Lambert} D.~L.,  1981, \mn@doi [\apj]
  {10.1086/159114}, \href {http://ads.bao.ac.cn/abs/1981ApJ...247.1052S} {247,
  1052}

\bibitem[\protect\citeauthoryear{{Tomkin}, {Lambert}, {Edvardsson},
  {Gustafsson}  \& {Nissen}}{{Tomkin} et~al.}{1989}]{Tomkin1989}
{Tomkin} J.,  {Lambert} D.~L.,  {Edvardsson} B.,  {Gustafsson} B.,   {Nissen}
  P.~E.,  1989, \aap, \href {http://ads.bao.ac.cn/abs/1989A%26A...219L..15T}
  {219, L15}

\bibitem[\protect\citeauthoryear{{Vennes}, {Christian}, {Mathioudakis}  \&
  {Doyle}}{{Vennes} et~al.}{1997}]{1997Vennes}
{Vennes} S.,  {Christian} D.~J.,  {Mathioudakis} M.,   {Doyle} J.~G.,  1997,
  \aap, \href {http://adsabs.harvard.edu/abs/1997A%26A...318L...9V} {318, L9}

\bibitem[\protect\citeauthoryear{{Wang}, {Liu}, {Zhao}  \& {Sato}}{{Wang}
  et~al.}{2011}]{WANG2011}
{Wang} L.,  {Liu} Y.,  {Zhao} G.,   {Sato} B.,  2011, \mn@doi [\pasj]
  {10.1093/pasj/63.5.1035}, \href
  {http://ads.bao.ac.cn/abs/2011PASJ...63.1035W} {63, 1035}

\bibitem[\protect\citeauthoryear{{Wright}, {Drake}, {Mamajek}  \&
  {Henry}}{{Wright} et~al.}{2011}]{2011Wright}
{Wright} N.~J.,  {Drake} J.~J.,  {Mamajek} E.~E.,   {Henry} G.~W.,  2011,
  \mn@doi [\apj] {10.1088/0004-637X/743/1/48}, \href
  {http://adsabs.harvard.edu/abs/2011ApJ...743...48W} {743, 48}

\bibitem[\protect\citeauthoryear{{Yi}, {Kim}  \& {Demarque}}{{Yi}
  et~al.}{2003}]{YonseiYale2003}
{Yi} S.~K.,  {Kim} Y.-C.,   {Demarque} P.,  2003, \mn@doi [\apjs]
  {10.1086/345101}, \href {http://ads.bao.ac.cn/abs/2003ApJS..144..259Y} {144,
  259}



\makeatother
\end{thebibliography}


\appendix
\section{Estimate of model abundances after pollution}


AGB stars loose mass, $\Delta$$\mathrm{M}_1$=$\Sigma$$\Delta$$\mathrm{M}_{tr}$, during TP-AGB phase. Every $\Delta$$\mathrm{M}_{tr}$ has different s-process abundances. For the sake of simplicity, $\Delta$$\mathrm{M}_1$ was assumed with a fixed abundance [X/Fe]$^{mid}$, which can be reasonable get from the models provided by \cite{2016Karakas}. At the end of AGB phase, a planetary nebula of mass $\Delta$$\mathrm{M}_2$ is ejected with a final element abundance [X/Fe]$^{fnl}$ \citep{Boffin1988}. Then, the total lost mass for the AGB star is $\mathrm{M}_{lost}$=$\Delta\mathrm{M}_1$ +$\Delta\mathrm{M}_2$.
The accretion efficiency $\alpha$ is estimated from $\mathrm{M}_{acc}$/$\mathrm{M}_{lost}$. $\mathrm{M}_{acc}$ means the accreted mass from the AGB donor. We assume $\Delta\mathrm{M}_1$ and $\Delta\mathrm{M}_2$ contribute the same portion in the accreted mass. Then, $\mathrm{M}_{acc}$=$\Delta\mathrm{M}_1^{'}$ +$\Delta\mathrm{M}_2^{'}$

The theoretical abundances after pollution for the Ba stars are estimated using the following,
$[X/Fe] = \log( 10^{[X/Fe]^{ini}}(1-r) + r(1-\gamma)10^{[X/Fe]^{mid}} + r \gamma10^{[X/Fe]^{fnl}})$
, [X/Fe]$^{ini}$ is the initial abundance of Ba stars before pollution, which we assumed same as the solar s-process abundance.
[X/Fe]$^{mid}$ is the abundance during the TP episodes on AGB, and we have taken a fixed value from models.
 $r$ is the `pollution factor' and expressed as $\mathrm{M}_{acc}$/$\mathrm{M}_{env}$.
$\gamma$ means the mass fraction of $\Delta\mathrm{M}_2^{'}$ with final abundances in the total accreted matter.
$\Delta\mathrm{M}_2^{'} = \gamma \mathrm{M}_{acc}$; $\Delta\mathrm{M}_1^{'} = (1- \gamma) \mathrm{M}_{acc}$. For RE~J0702+129, we have assumed that $\Delta\mathrm{M}_1$ and $\Delta\mathrm{M}_2$ contribute the same portion in the accreted mass, then, $\gamma$ = $\Delta\mathrm{M}_2^{'}/\mathrm{M}_{acc}$ = $\Delta\mathrm{M}_2$/$\mathrm{M}_{lost}$=0.80/(0.08+0.80)=0.91. For BD+80$^\circ$670, $\gamma$ = 1.60/(1.18+1.60)=0.58.

\begin{table*}
  \centering
  \caption{Adopted AGB models and their parameters}
    \begin{tabular}{|l|c|c|c|c|c|c|c|c|c|c|}
    \toprule
          & \multicolumn{5}{c|}{M=1.5 ${M_\odot}$}            & \multicolumn{5}{c|}{M=3.5 ${M_\odot}$} \\
    \midrule
    & \multicolumn{1}{l|}{$[X/Fe]^{fnl}$} & \multicolumn{1}{l|}{$[X/Fe]^{mid}$} & \multicolumn{1}{l|}{r=0.08} & \multicolumn{1}{l|}{r=0.2} & \multicolumn{1}{l|}{r=0.4} & \multicolumn{1}{l|}{$[X/Fe]^{fnl}$} & \multicolumn{1}{l|}{$[X/Fe]^{mid}$} & \multicolumn{1}{l|}{r=0.06} & \multicolumn{1}{l|}{r=0.11} & \multicolumn{1}{l|}{r=0.2} \\
    \midrule
    $[Sr/Fe]$   & 0.60  & 0.00  & 0.09  & 0.19  & 0.32  & 1.27  & 1.14  & 0.29  & 0.43  & 0.61  \\
    \midrule
   $[ Y/Fe]$    & 0.68  & 0.00  & 0.11  & 0.23  & 0.38  & 1.34  & 1.21  & 0.33  & 0.49  & 0.68  \\
    \midrule
    $[Zr/Fe]$    & 0.73  & 0.00  & 0.12  & 0.25  & 0.41  & 1.33  & 1.21  & 0.32  & 0.48  & 0.67  \\
    \midrule
    $[Ba/Fe]$    & 1.00  & 0.00  & 0.22  & 0.42  & 0.63  & 1.26  & 1.18  & 0.29  & 0.44  & 0.62  \\
    \midrule
    $[La/Fe]$   & 0.96  & 0.00  & 0.20  & 0.39  & 0.60  & 1.17  & 1.09  & 0.25  & 0.38  & 0.55  \\
    \midrule
    $[Ce/Fe]$    & 0.95  & 0.00  & 0.19  & 0.38  & 0.59  & 1.14  & 1.06  & 0.23  & 0.36  & 0.53  \\
    \midrule
    $[Nd/Fe]$    & 0.80  & 0.00  & 0.14  & 0.29  & 0.46  & 0.96  & 0.90  & 0.16  & 0.27  & 0.40  \\
    \bottomrule
    \end{tabular}%
  \label{tab:addlabel}%
\end{table*}%

\begin{table*}
  \centering
  \caption{Estimated model parameters for our sample stars}
    \begin{tabular}{lcccccccccr}
    \toprule
          & \multicolumn{1}{l}{M/${M_\odot}$} & \multicolumn{1}{l}{$M^{env}$/${M_\odot}$} & \multicolumn{2}{c}{$M_{lost}$/${M_\odot}$} &  \multicolumn{1}{c}{$\gamma$} & \multicolumn{1}{c}{$M_{acc}$/${M_\odot}$} & accretion efficiency     & pollution factor&
 \\

          &       &       & \multicolumn{1}{l}{$\Delta$M1} & \multicolumn{1}{l}{$\Delta$M2} &               &   &  &  \\
    \midrule
    RE~J0702+129 & 0.93  & 0.0076 & 0.08  & 0.80 & 0.91  & 0.0006-0.003      & 0.07\% - 0.35\% &  0.08 - 0.40  &\\
     \midrule
    {\bf BD+80$^\circ$670} & 1.05  & 0.0451 & 1.18 & 1.60 & 0.58 &0.0027-0.005        & 0.10\% - 0.18\% & 0.06 - 0.11 &\\
     \bottomrule
    \end{tabular}%
  \label{tab:addlabel}%
\end{table*}%


\bsp	
\label{lastpage}

\end{document}